\documentclass[prd,onecolumn]{revtex4}
\usepackage{dcolumn}
\usepackage{array}
\usepackage{longtable}
\usepackage{multirow}
\usepackage{graphicx}
\usepackage{amssymb}
\usepackage{bm}
\usepackage{hyperref}
\usepackage{epstopdf}
\usepackage{color}
\usepackage{mathrsfs}
\usepackage{amsmath,amssymb,amsthm}
\begin{document}
\title{Searching for the Evidence of Dynamical Dark Energy}

\author{Deng Wang$^{1, 2}$}
\email{cstar@mail.nankai.edu.cn, cstar@sjtu.edu.cn}
\author{Wei Zhang$^{3}$}
\email{cosmoszhang@mail.nankai.edu.cn}
\author{Xin-He Meng$^{3}$}
\email{xhm@nankai.edu.cn}

\affiliation{
$^1${Department of Astronomy, School of Physics and Astronomy, Shanghai Jiao Tong University, Shanghai 200240, China}\\
$^2${Theoretical Physics Division, Chern Institute of Mathematics, Nankai University, Tianjin 300071, China}\\
$^3${Department of Physics, Nankai University, Tianjin 300071, China}}

\begin{abstract}
In the statistical framework of model-independent Gaussian processes (GP), we search for the evidence of dynamical dark energy (DDE) using the `` Joint Light-curve Analysis '' (JLA) Type Ia supernovae (SNe Ia) sample, the 30 latest cosmic chronometer data points (H(z)), Planck's shift parameter from cosmic microwave background (CMB) anisotropies, the 156 latest HII galaxy measurements and 79 calibrated gamma-ray bursts (GRBs). We find that the joint constraint from JLA + H(z) + CMB + HII + GRB supports the global measurement of $H_0$ by Planck collaboration very much in the low redshift range $z\in[0, 0.76]$ at the $2\sigma$ confidence level (C.L.), gives a cosmological constant crossing (quintom-like) equation of state (EoS) of DE at the $2\sigma$ C.L. and implies that the evolution of the late-time Universe may be actually dominated by the DDE.
\end{abstract}
\maketitle
\section{Introduction}
With gradually accumulating data, modern cosmological observations including SNe Ia, CMB anisotropies, baryonic acoustic oscillations (BAO), observational Hubble parameter, the abundance of galaxy clusters (AGC), strong and weak gravitational lensing (SGL/WGL), etc., have strongly indicated that our Universe is undergoing a phase of accelerated expansion \cite{1,2,3,4}. To explain this accelerated mechanism, theorists have proposed an exotic and negative pressure fluid dubbed dark energy (DE). The simplest model to characterize the phenomena is the so-called $\Lambda$-cold-dark-matter ($\Lambda$CDM) model, which is described by the EoS of DE $\omega=-1$. In the past two decades, this model has been proved to be very consistent with most of the observed data, containing the SNe Ia, BAO and so on. Nonetheless, the newest results by the Planck satellite imply that there still exist some anomalies which are non-compatible with the predictions of the $\Lambda$CDM model \cite{5}, including the anomalies of the observed $H(z)$ data and the amplitude of fluctuation spectrum. Apart from these anomalies, this scenario also faces two unsolved puzzles, i.e., the well-known fine-tuning and coincidence problems \cite{6}: The former indicates the theoretical value for the vacuum energy density are far greater than its observed value ($\rho^{the}_{vac}\gg\rho^{obs}_{vac}$), namely the so-called 120-orders-of-magnitude discrepancy that makes the vacuum explanation very confusing; while the latter is why the energy densities of the dark matter (DM) and DE are of the same order at the present epoch, since their energy densities are so different from each other during the evolutional process of the Universe. Based on these concerns, a question naturally comes into being: Is the DE actually dynamical ($\omega\neq-1$) or dominated by a cosmological $\Lambda$ term ?

To address this issue, usually speaking, one should utilize the parametric or nonparametric methods to reconstruct the evolution of $\omega(z)$ ($z$ being the redshift) from the observed data \cite{7,8,9,10,11,12,13,14,15,16,17,18}: The former cases are implemented by either assuming an ad hoc parametrization form of the EoS, energy density, or pressure of DE, or developing a  concrete cosmological model based on some physical mechanism ; while the latter cases reconstruct $\omega(z)$ starting directly from data without assuming a specific form of $\omega(z)$, but need more statistical inputs. With the explosion of data in recent years, using the non-parametric methods to extract the information hiding in data has inspired a new fashion. In terms of reconstructions of $\omega(z)$, here we review two main methods as follows:

$\star$ Binning method \cite{add}: This method bins $\omega$ in $z$ and fits the amplitudes to observed data by assuming $\omega(z)$ is constant within each bin. In practice, the keynote of this method is choosing the proper bin number $N$ for the corresponding data. When $N$ is very large, the uncertainties of binned $\omega$'s are very large and highly correlated, representing the flat directions in the likelihood function and a substantially slow convergence of Monte Carlo Markov Chains (MCMC). Conversely, when $N$ is very small, the convergence can be reached fast, but the coarse binning leads to unphysical discrete structures. This can be attributed to the prior assumption that there is a perfect correlation of $\omega(z)$ within each bin and no correlation among different bins.

$\star$ Principal component analysis (PCA) \cite{19}: This method is a useful statistical tool to compress data, and provide a compatible framework for forecasting and comparing the information content of future surveys \cite{20}. It has been applied in reconstructing the EoS of DE $\omega(z)$ \cite{add,22}. The optimal basis of PCA for a cosmological quantity can be obtained by implementing a Fisher forecast to determine the eigenmodes of the covariance matrix based on a specific reference model, and then fitting data to the coefficients of the principal components through $\chi^2$ minimization. Since ignoring the contributions of small enough eigenvalues and implementing a truncation, a small part of data information must be lost. Additionally, zeroing the noisy modes also introduces a hidden prior on the smoothness of $\omega(z)$ that is difficult to quantify and interpret.

In light of the problems that these two methods face when reconstructing $\omega(z)$, T. Holsclaw et al. proposed a powerful nonparametric method based on GP modeling and MCMC sampling \cite{23}. It possesses the following several advantages: (i) it avoids artificial biases due to restricted parametric assumptions for $\omega(z)$; (ii) it does not lose information of data by smoothing it; (iii) it can control the errors effectively, rather than introduce arbitrariness by using a certain number of bins to describe data or truncating information using a restricted set of optimal basis functions to describe data. By analyzing the Constitution SNe Ia data set \cite{24}, they found that the reconstructed $\omega(z)$ is consistent with the $\Lambda$CDM model at the $1\sigma$ C.L.. After that, using the updated GP method and the Union 2.1 data set, M. Seikel et al. concluded that there still exists a high degeneracy between the $\Lambda$CDM model and DDE models at low redshifts \cite{25}. Most recently, following this logical line, we modify the available online package GaPP (Gaussian processes in Python) invented by M. Seikel et al. \cite{25}, add the 30 latest cosmic chronometer data points and Planck's shift parameter, and find that the GP reconstructions of $\omega(z)$ are still consistent with the $\Lambda$CDM model at the $2\sigma$ C.L. \cite{Deng}. This gives a underlying possibility of the existence of DDE if one uses more high-quality data to reconstruct the EoS of DE $\omega(z)$. In the present study, we continue exploring the possible deviations from the the $\Lambda$CDM model by using the largest JLA SNe Ia sample \cite{26}, cosmic chronometer data, CMB observation, the latest HII galaxy measurements and calibrated high-$z$ gamma-ray burst (GRB) data.

This study is organized in the following manner. In Sec. II, we review briefly on the GP methodology. In Sec. III, we describe the observational data used in this analysis, containing SNe Ia, H(z), CMB, HII galaxies and GRB. In Sec. IV, we exhibit the results of the GP reconstructions. The discussions and conclusions are presented in the final section.

\section{Methodology}
As described in \cite{23,25}, the model-independent GP is a fully Bayesian approach for smoothing data, and can reconstruct directly a function from the observational data without assuming a specific model or choosing a parametrization form for the underlying function. As a consequence, it has been widely applied in modern observational cosmology such as, investigating the expansion dynamics of the universe \cite{23,25,27}, the distance duality relation \cite{28}, the cosmography \cite{29}, the null test of the cosmological constant \cite{30}, the determination of the interaction between dark energy and dark matter \cite{31}, dodging the matter degeneracy to determine the dynamics of dark energy \cite{32}, the slowing down of cosmic acceleration \cite{33,34}, dodging the cosmic curvature to probe the constancy of the speed of light \cite{35}, and so forth.

For a Friedmann-Robertson-Walker (FRW) Universe in the framework of general relativity (GR), the luminosity distance $d_L(z)$ is expressed as
\begin{equation}
d_L(z)=\frac{c(1+z)}{H_0\sqrt{|\Omega_{k0}|}}sinn\left(\sqrt{|\Omega_{k0}|}\int^{z}_{0}\frac{dz'}{E(z')}\right), \label{1}
\end{equation}
where the dimensionless Hubble parameter $E(z)=H(z)/H_0$, the present-day cosmic curvature $\Omega_{k0}=-Kc^2/(a_0H_0^2)$, and for $sinn(x)= sin(x), x, sinh(x)$, $K=1, 0, -1$ , which corresponds to a closed, flat and open Universe, respectively. Using the normalized comoving distance $D(z)=(H_0/c)(1+z)^{-1}d_L(z)$, the EoS of DE is written as
\begin{equation}
\omega(z)=\frac{2(1+z)(1+\Omega_{k0})D''-[(1+z)^2\Omega_{k0}D'^2-3(1+\Omega_{k0}D^2)+2(1+z)\Omega_{k0}DD']D'}{3D'\{(1+z)^2[\Omega_{k0}+(1+z)\Omega_{m0}]D'^2-(1+\Omega_{k0}D^2)\}}, \label{2}
\end{equation}
where $\Omega_{m0}$ is the present-day matter density ratio parameter and the prime denotes the derivative with respect to (w.r.t.) $z$. Assuming $\Omega_{m0}=0.308\pm0.012$ from the recent Planck-2015 results \cite{5}, we just consider the possibility of the existence of DDE in a spatially flat FRW Universe ($\Omega_{k0}=0$) throughout this work, Eq. (\ref{2}) can be rewritten as
\begin{equation}
\omega(z)=\frac{2(1+z)D''+3D'}{3D'[(1+z)^3\Omega_{m0}D'^2-1]}. \label{3}
\end{equation}

We take the public package GaPP to implement the reconstructions.
Generally speaking, the GP is a generalization of a Gaussian distribution, which is the distribution of a random variable, and exhibits a distribution over functions.
At each reconstructed point $z$, the reconstructed function $f(z)$ is a Gaussian distribution with a mean value and Gaussian error. The key point of the GP is a covariance function $k(z,\tilde{z})$ which correlates the function $f(z)$ at different reconstructed points. More precisely, the covariance function $k(z,\tilde{z})$ depends only on two hyperparameters $l$ and $\sigma_f$, which characterize the coherent scale of the correlation in $x$-direction and typical change in the $y$-direction, respectively. In general, the choice is the squared exponential covariance function $k(z,\tilde{z})=\sigma_f^2 \mathrm{exp}[-|z-\tilde{z}|^2/(2l^2)]$. However, M. Seikel et al. \cite{36} have demonstrated that the Mat\'{e}rn ($\nu=9/2$) covariance function is a better choice to carry out the reconstructions. Therefore, in the following analysis, we choose the Mat\'{e}rn ($\nu=9/2$) covariance function:
\begin{equation}
k(z,\tilde{z})=\sigma_f^2 \mathrm{exp}(-\frac{3|z-\tilde{z}|}{l})\times[1+\frac{3|z-\tilde{z}|}{l}+\frac{27(z-\tilde{z})^2}{7l^2}+\frac{18|z-\tilde{z}|^3}{7l^3}+\frac{27(z-\tilde{z})^4}{35l^2}]. \label{4}
\end{equation}
This indefinitely differentiable function is very useful to reconstruct the derivatives of a specific function.
\section{Data}
In this section, we introduce the observational data used in the GP reconstructions including SNe Ia, H(z), CMB, HII galaxy and GRB observations.

The observations of SNe Ia provide a useful tool to probe the dark dynamics and expansion history of the Universe. It is well known that the absolute magnitudes of all the SNe Ia are considered to be the same, since all the SNe Ia almost explode at the same mass ($M\approx-19.3\pm0.3$). For this reason, SNe Ia can theoretically act as the standard candles to constrain different cosmological models. In this situation, we adopt the JLA sample containing 740 SNe Ia data points, which covers the redshift range $z \in [0.01, 1.3]$ \cite{26}. The JLA sample can generally be divided into four classes: (i) 118 low-$z$ SNe in the range $z \in [0, 0.1]$ from \cite{24,37,38,39,40,41}; (ii) 374 SNe in the range $z \in [0.3, 0.4]$ from the Sloan Digital Sky Survey (SDSS) SNe search \cite{42}; (iii) 239 SNe in the range $z \in [0.1, 1.1]$ from the Supernova Legacy Survey (SNLS) project \cite{43}; (iv) 9 high-$z$ SNe in the range $z \in [0.8, 1.3]$ from the Hubble Space Telescope (HST) \cite{56}. As noted in \cite{25}, we transform the distance modulus $m-M$ of JLA data to $D$ by utilizing the following formula:
\begin{equation}
m-M-25+5\log(\frac{H_0}{c})=5\lg[(1+z)D], \label{5}
\end{equation}
where `` log '' denotes the logarithm to base 10. Throughout the reconstruction process, we set the initial conditions $D(z=0)$ and $D'(z=0)=1$. From Eq. (\ref{5}), one can easily find that the values of $D$ only depend on a combination of the absolute magnitude $M$ and the Hubble constant $H_0$.

To date, there are two main methods to obtain H(z) data, i.e., the radial BAO and galaxy differential age methods. As usual, to obtain H(z) values from the radial BAO method, one needs to model the redshift space distortions (RSD) and assume an acoustic scale, both of which require the assumption of a particular cosmological model. Thus, the H(z) data obtained from the radial BAO method is actually model-dependent. To be different, the H(z) data from the galaxy differential age method is based on the direct model-independent observations. Hence, we still use the 30 latest cosmic chronometer data points compiled in our previous work (see Table. I in \cite{Deng}): 5 from \cite{44}; 1 from \cite{45}; 8 from \cite{46}; 7 from \cite{47}; 5 from \cite{48}; 2 from \cite{49}; 2 from \cite{50}. As before, we implement the reconstructions by transforming $H(z)$ to $D'$.

As done in \cite{Deng}, we continue using the CMB shift parameter $\mathcal{R}=1.7488\pm0.0074$ from the recent Planck's release \cite{3} to provide an extremely high-$z$ constraint (see Figs. 3-4 in \cite{Deng}). To be more precise, one can obtain practically the improved constraint by transforming $\mathcal{R}=\sqrt{\Omega_{m0}}\int^{z_c}_0\frac{dz'}{E(z')}$ to $D$, where where $z_c=1089.0$ is the redshift of recombination.

As an important supplement for the SNe Ia observations, we also use 156 HII galaxy measurements to implement our GP reconstructions: (i) 24 Giant Extragalactic HII Regions (GEHR) at redshifts $z\leqslant0.01$ \cite{51}; (ii) 107 low-$z$ HII galaxies \cite{52}; (iii) 25 high-$z$ HII galaxy measurements include 19 high-$z$ objects(1 from \cite{53}, 6 from \cite{54} and 12 from \cite{55}) and 6 high-$z$ star-forming galaxies in the redshift range $z\in[0.64, 2.33]$ obtained via the X-Shooter spectrography at the Cassegrain focus of the European Southern Observatory Very Large Telescope (ESO-VLT) \cite{56}. In \cite{57}, it has been verified that
for GEHR and HII galaxies, the $L(H\beta)-\sigma$ relation can be applied into measuring the distance, and it can be written as
\begin{equation}
\mathrm{log}L(H\beta)=(5.05\pm0.097)\mathrm{log}\sigma(H\beta)+(33.11\pm0.145), \label{6}
\end{equation}
where $L(H\beta)$ and $\sigma(H\beta)$ represent the Balmer emission line luminosity for these objects and the velocity dispersion of the young star-forming cluster from the measurements of the line width, respectively. Subsequently, the corresponding observed distance modulus is expressed as
\begin{equation}
\mu_{obs}=m-M=2.5\mathrm{log}L(H\beta)-2.5\mathrm{log}f(H\beta)-100.95, \label{7}
\end{equation}
where $f(H\beta)$ is the measured flux in the $H\beta$ line. Furthermore, for the purpose to use HII galaxy data, we obtain the $1\sigma$ statistical error of $\mu_{obs}$ by error propagations, and transform the distance modulus $m-M$ of HII galaxy data to $D$ by using Eq. (\ref{5}).

The GRB observations, which are among the most powerful sources in the Universe, are another useful high-$z$ supplement for the SNe Ia observations to carry out the GP reconstructions. The high energy photons of GRBs in the gamma-ray band are almost immune to dust extinction, and consequently they can be observed up to redshift $z\sim8-9$ \cite{58,59}, which goes beyond the redshift range of observed SNe Ia ($z<2$). Therefore, we might use the GRB probe to explore the early universe and provide an effective high-$z$ constraint on the EoS of DE. In this analysis, we adopt the 79 GRBs covering the redshift range $z\in[1.44, 8.1]$ obtained by J. Liu et al. \cite{60}, who utilized the Union 2.1 SNe Ia data set to calibrate 138 long Swift GRBs based on the model-independent $P\acute{a}de$ method. As done for SNe Ia and HII galaxies, we also transform the distance modulus $m-M$ of GRBs to $D$ by using Eq. (\ref{5}).

\begin{figure}
\centering
\includegraphics[scale=0.33]{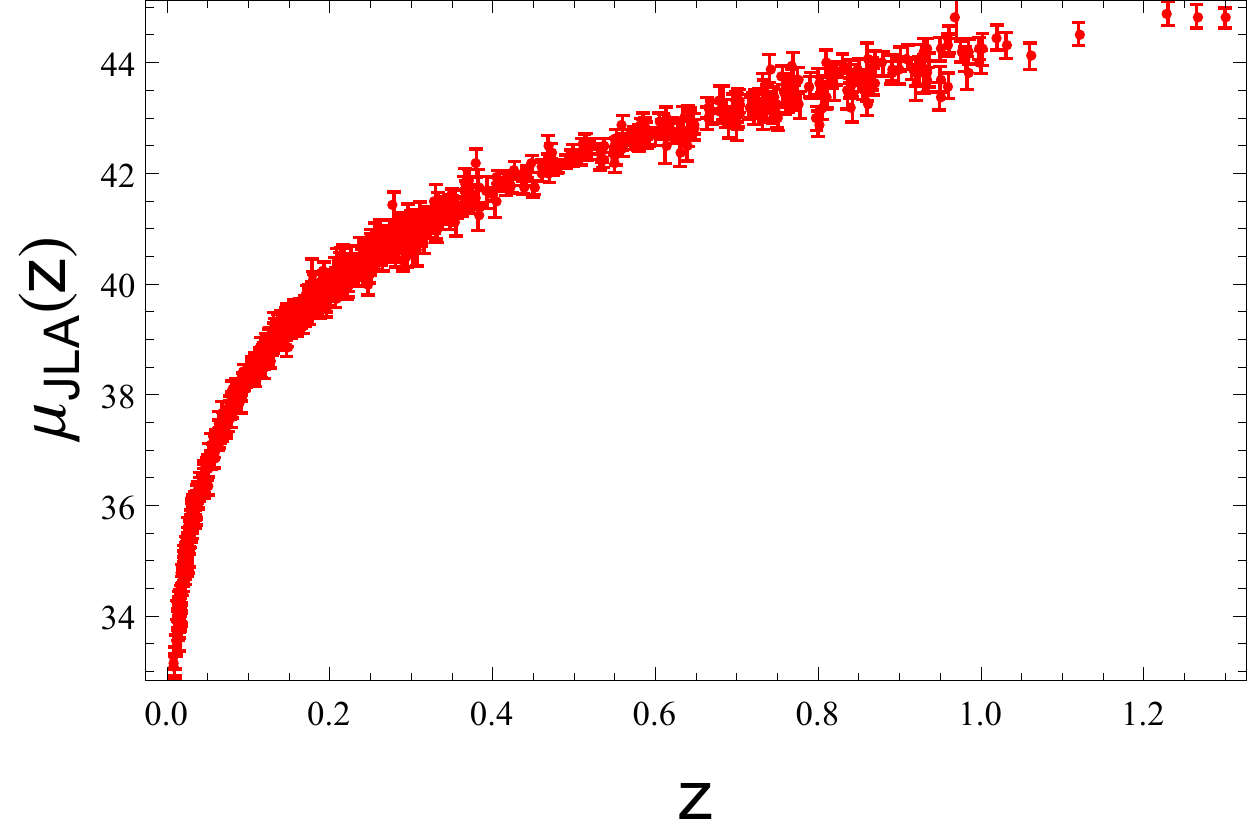}
\includegraphics[scale=0.33]{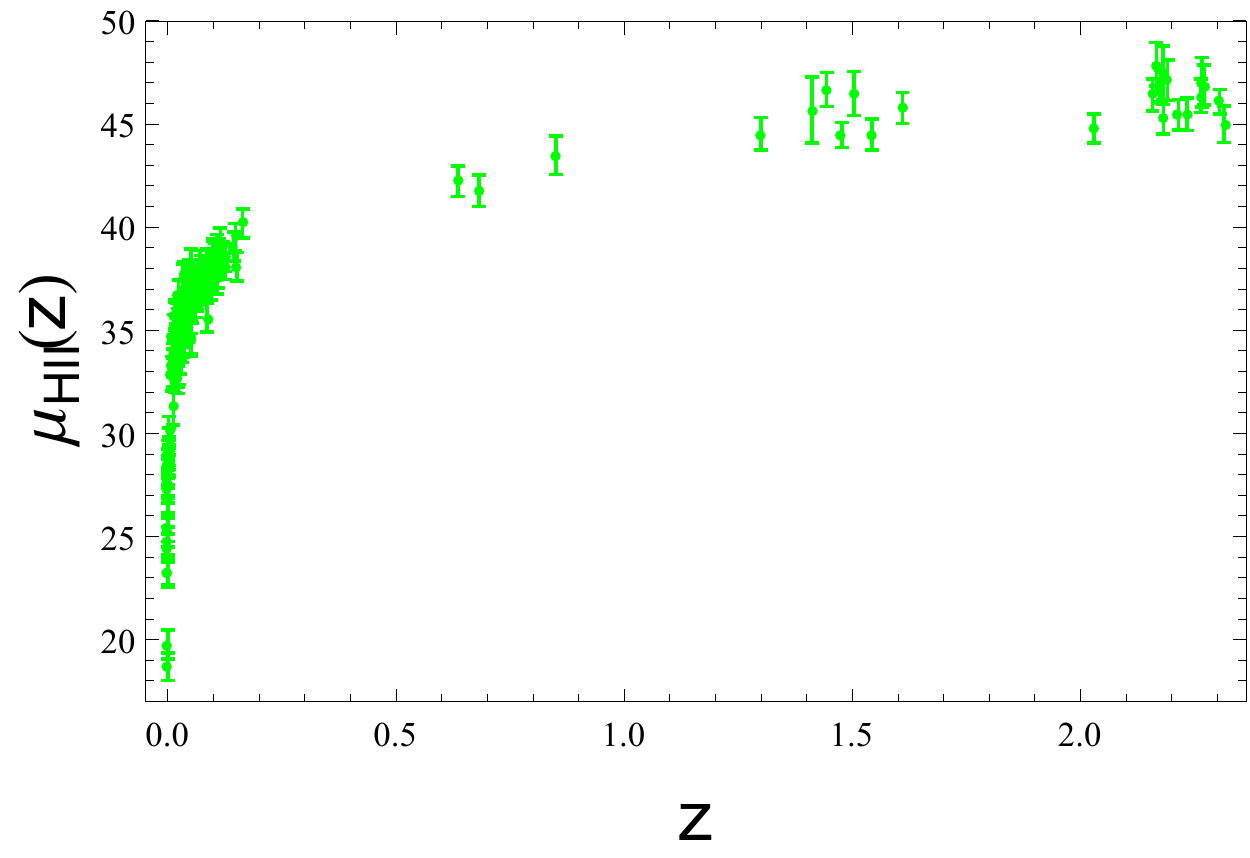}
\includegraphics[scale=0.33]{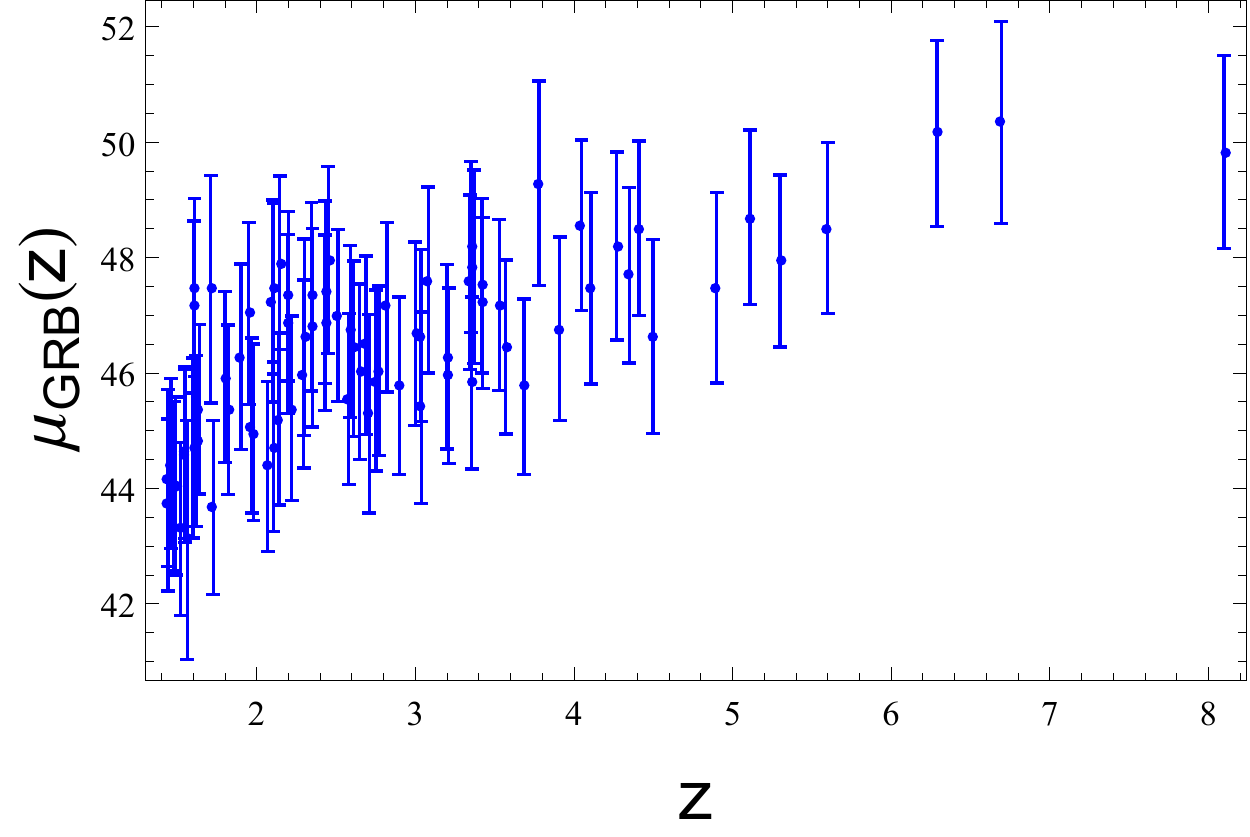}
\includegraphics[scale=0.34]{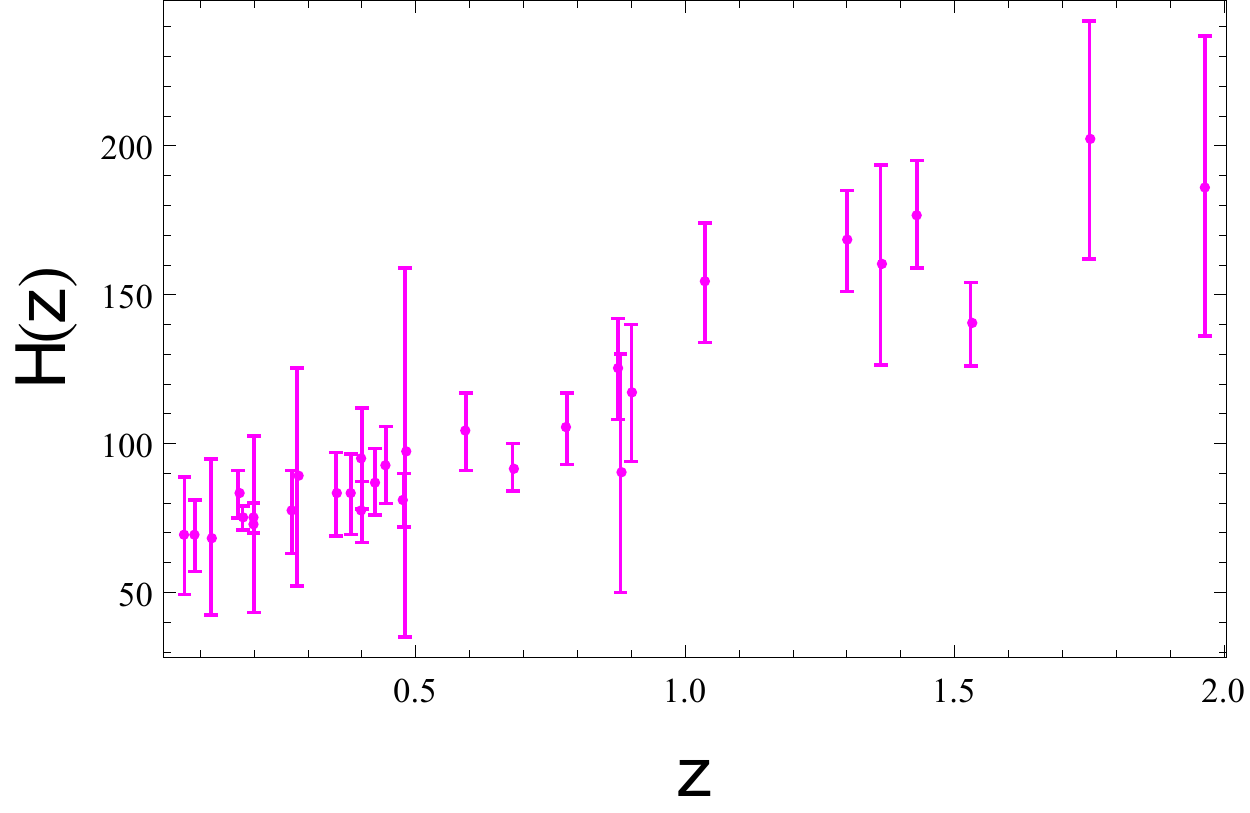}
\caption{From left to right: the relation between the distance modulus of SNe Ia and redshift $z$, the relation between the distance modulus of HII galaxies and redshift $z$, the relation between the distance modulus of GRBs and redshift $z$ as well as the relation between the Hubble parameter and redshift $z$. The dots with errors bar correspond to the observed data.}\label{f1}
\end{figure}

To exhibit the relations between the reconstructed $D$ and the above-mentioned data more clearly, we update the `` relations '' in \cite{Deng} as follows
$$ relations\Longrightarrow\left\{
\begin{aligned}
D & \Longleftarrow   m-M                                                         & \Longleftarrow \mathrm{JLA + HII + GRB}                                                       \\
D & \Longleftarrow  \mathcal{R}=\sqrt{\Omega_{m0}}\int^{z_c}_0\frac{dz'}{E(z')}  & \Longleftarrow \mathrm{CMB}            \\
D' & \Longleftarrow  \frac{H_0}{H(z)}                                            & \Longleftarrow \mathrm{H(z)}  \\
\end{aligned}
\right.
$$
Furthermore, in Fig. \ref{f1}, we also plot for the observed data used in this analysis.
In total, different from \cite{Deng}, we use the JLA SNe Ia data set, which has larger sample size and higher quality than the Union 2.1 sample, the latest HII galaxy measurements and complementary high-$z$ GRB probes to carry out the GP reconstructions.

\section{Results}
In this section, we combine five cosmological probes from different physical scales. i.e., SNe Ia, H(z), CMB, HII galaxies and GRB to explore the possibility of the existence of DDE. Meanwhile, we also try to report an intermediate result for the recent $H_0$ tension, namely the local value $73.24\pm1.74$ km s$^{-1}$ Mpc$^{-1}$ measured by A. Riess et al. \cite{61} (hereafter R16) is $3.4\sigma$ higher than the global value $66.93\pm0.62$ km s$^{-1}$ Mpc$^{-1}$ predicted by Planck collaboration\cite{5} (hereafter P16). Since we have pointed out that the value of variable $H_0$ affects the reconstructions of the EoS of DE by affecting obviously those of $D(z), D'(z)$ and $D''(z)$ in \cite{Deng}, we choose the representative values of $H_0$ obtained by two groups R16 and P16 to implement the reconstruction processes.
\begin{figure}
\centering
\includegraphics[scale=0.4]{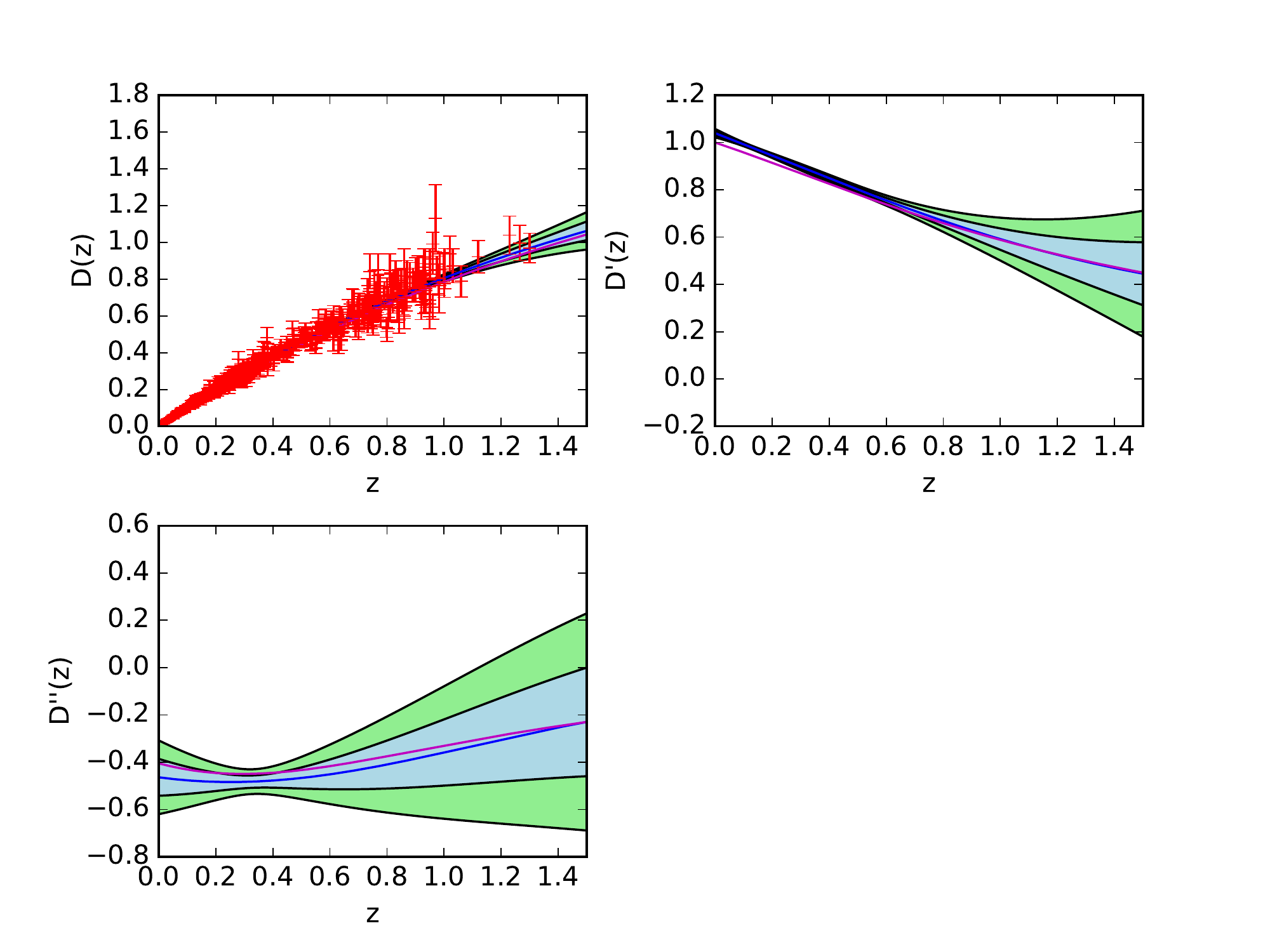}
\includegraphics[scale=0.4]{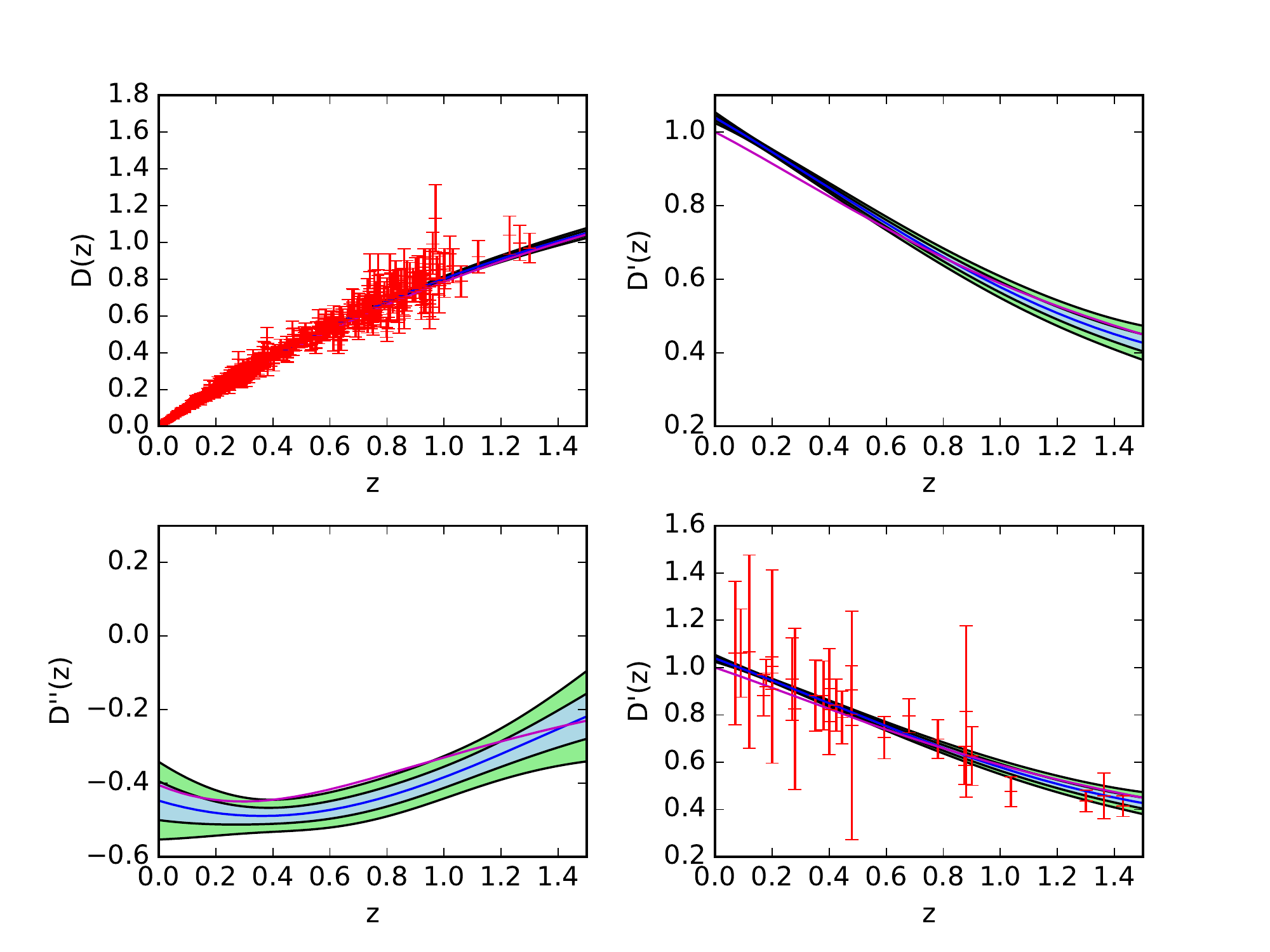}
\includegraphics[scale=0.4]{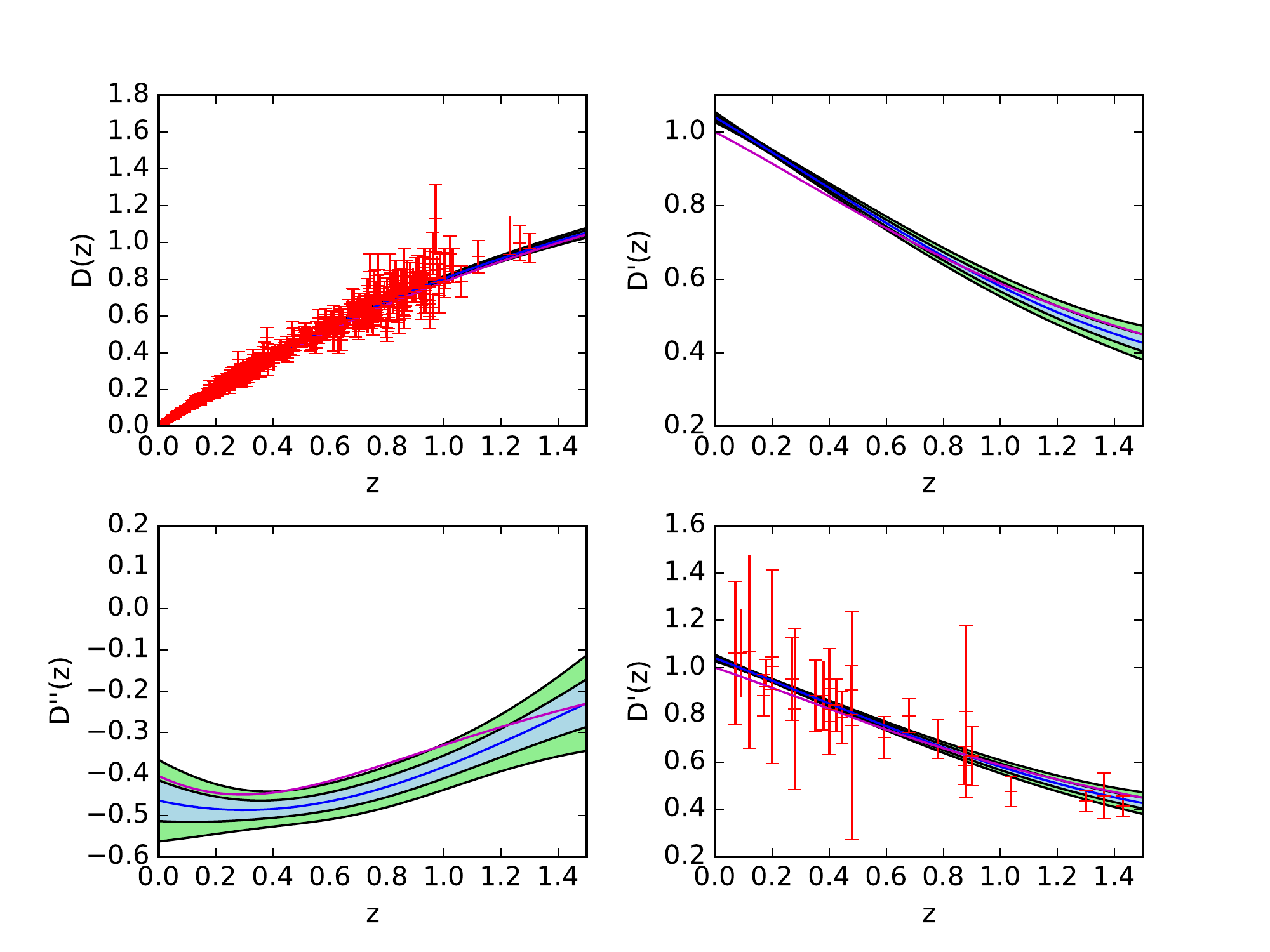}
\includegraphics[scale=0.4]{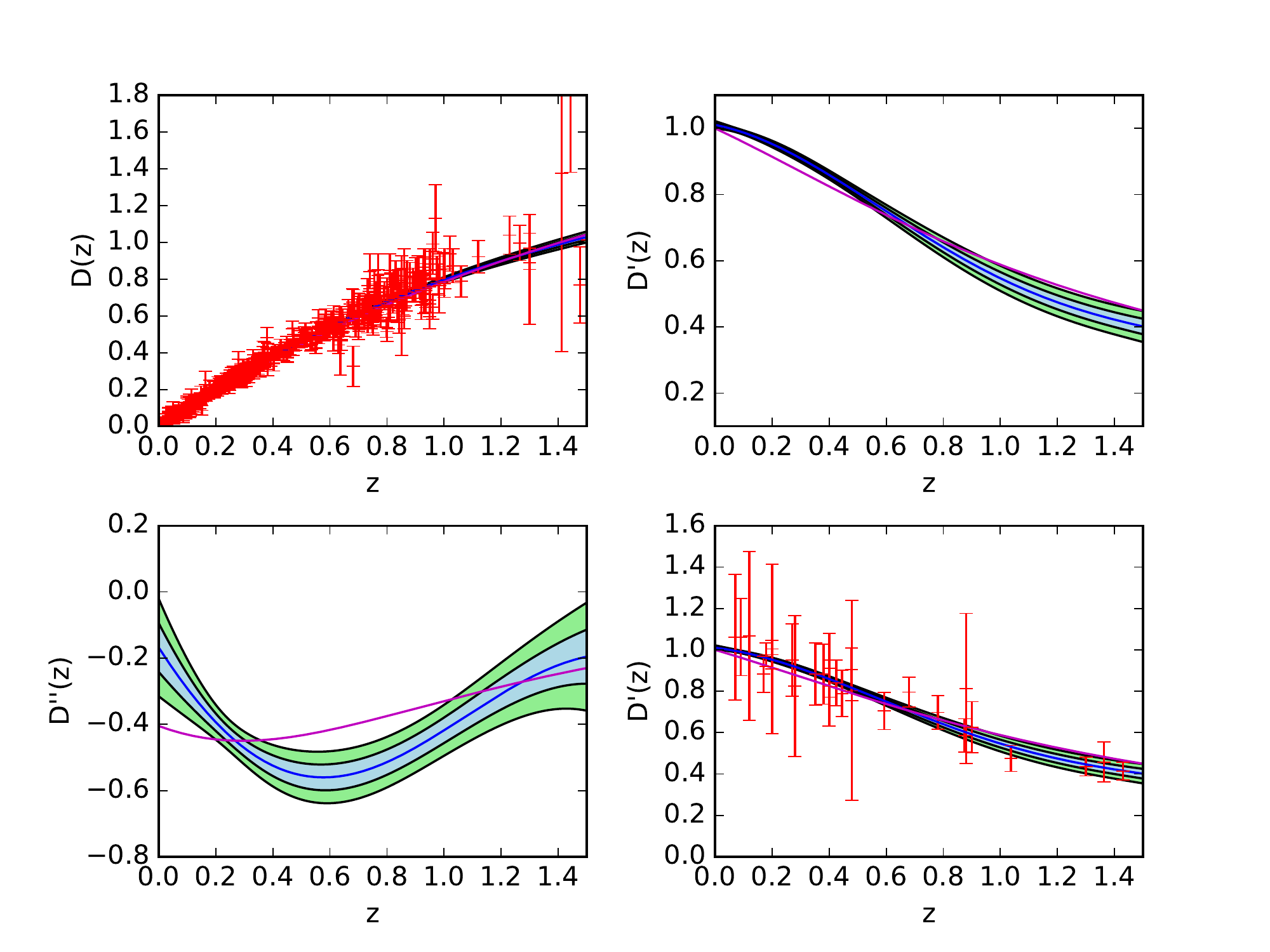}
\includegraphics[scale=0.4]{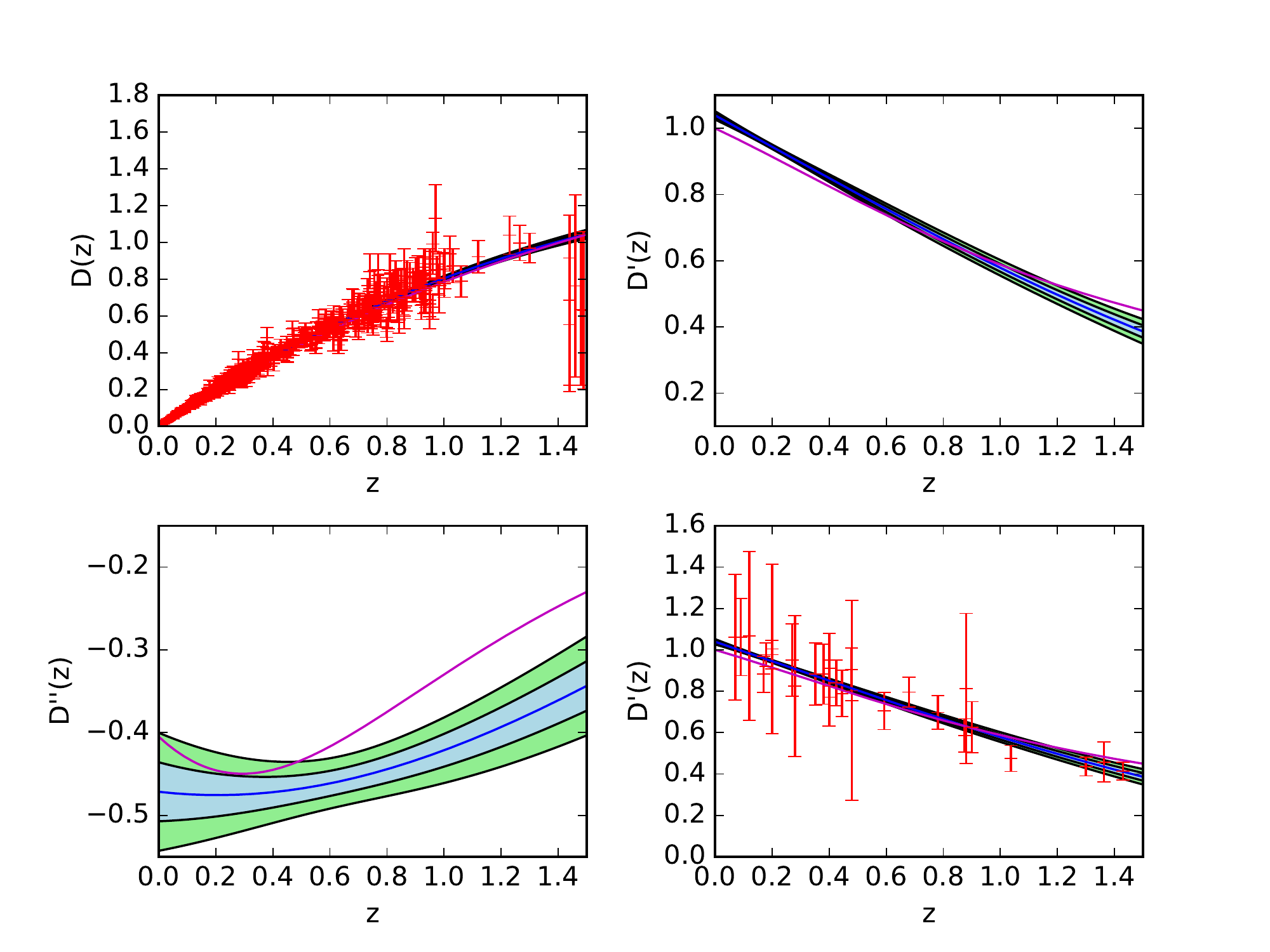}
\includegraphics[scale=0.4]{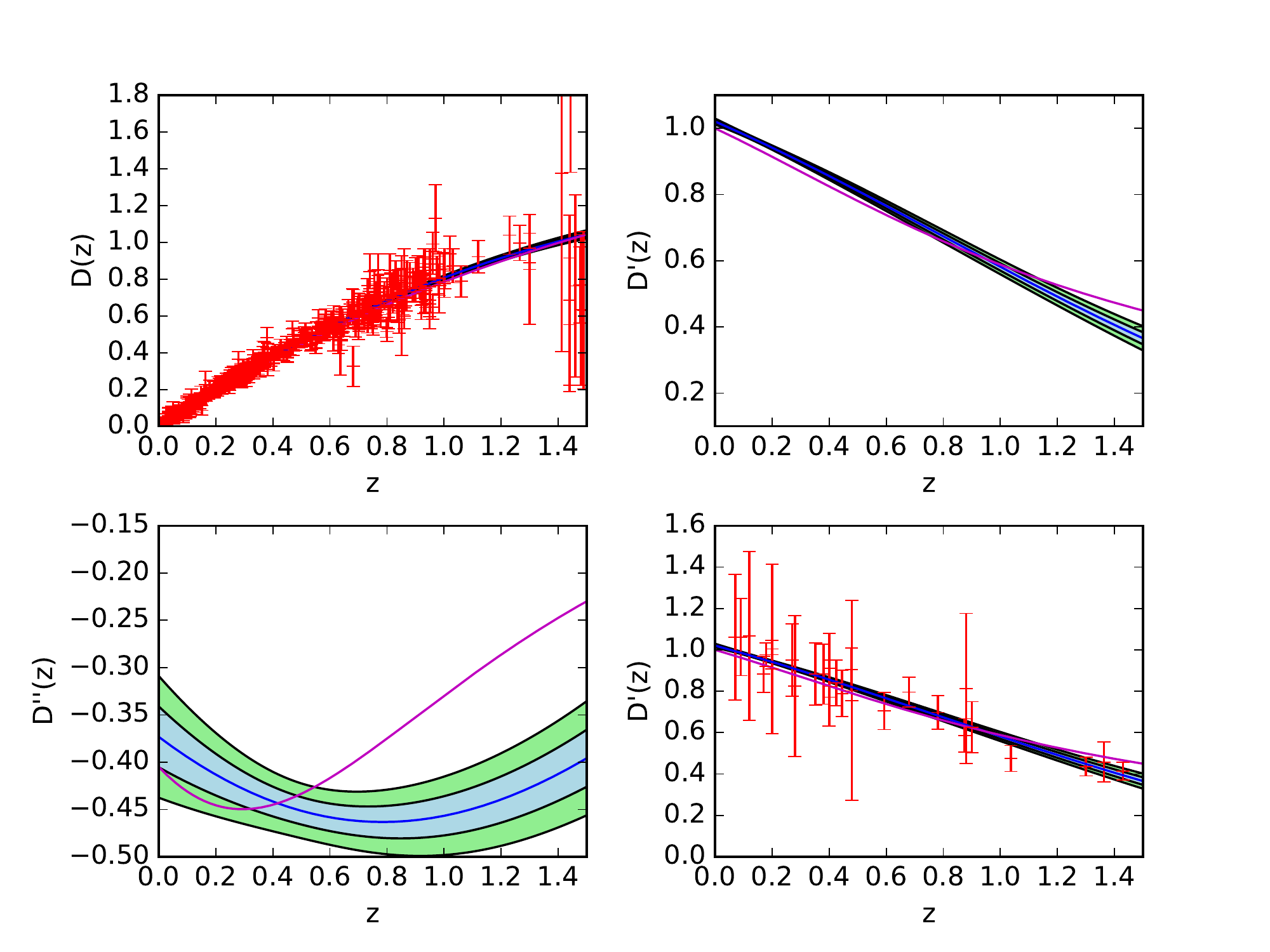}
\caption{The GP reconstructions of $D(z), D'(z)$ and $D''(z)$ using different observations. The upper left panel, upper right panel, middle left panel, middle right panel, lower left panel and lower right panel correspond to JLA, JLA + H(z), JLA + H(z) + CMB, JLA + H(z) + CMB + HII, JLA + H(z) + CMB + GRB and JLA + H(z) + CMB + HII + GRB, respectively. The observed distance modulus data points are shown in the upper left small panel of each panel.  Except for the upper left panel, the 30 latest $H(z)$ cosmic chronometer data points are shown in the lower right small panel of each panel. The blue and magenta lines represents the underlying true model (the mean value of reconstructions) and the $\Lambda$CDM model, respectively.  We have assumed $\Omega_{m0}=0.308\pm0.012$, $\Omega_{k0}=0$ and $73.24\pm1.74$ km s$^{-1}$ Mpc$^{-1}$.}\label{f2}
\end{figure}
\begin{figure}
\centering
\includegraphics[scale=0.23]{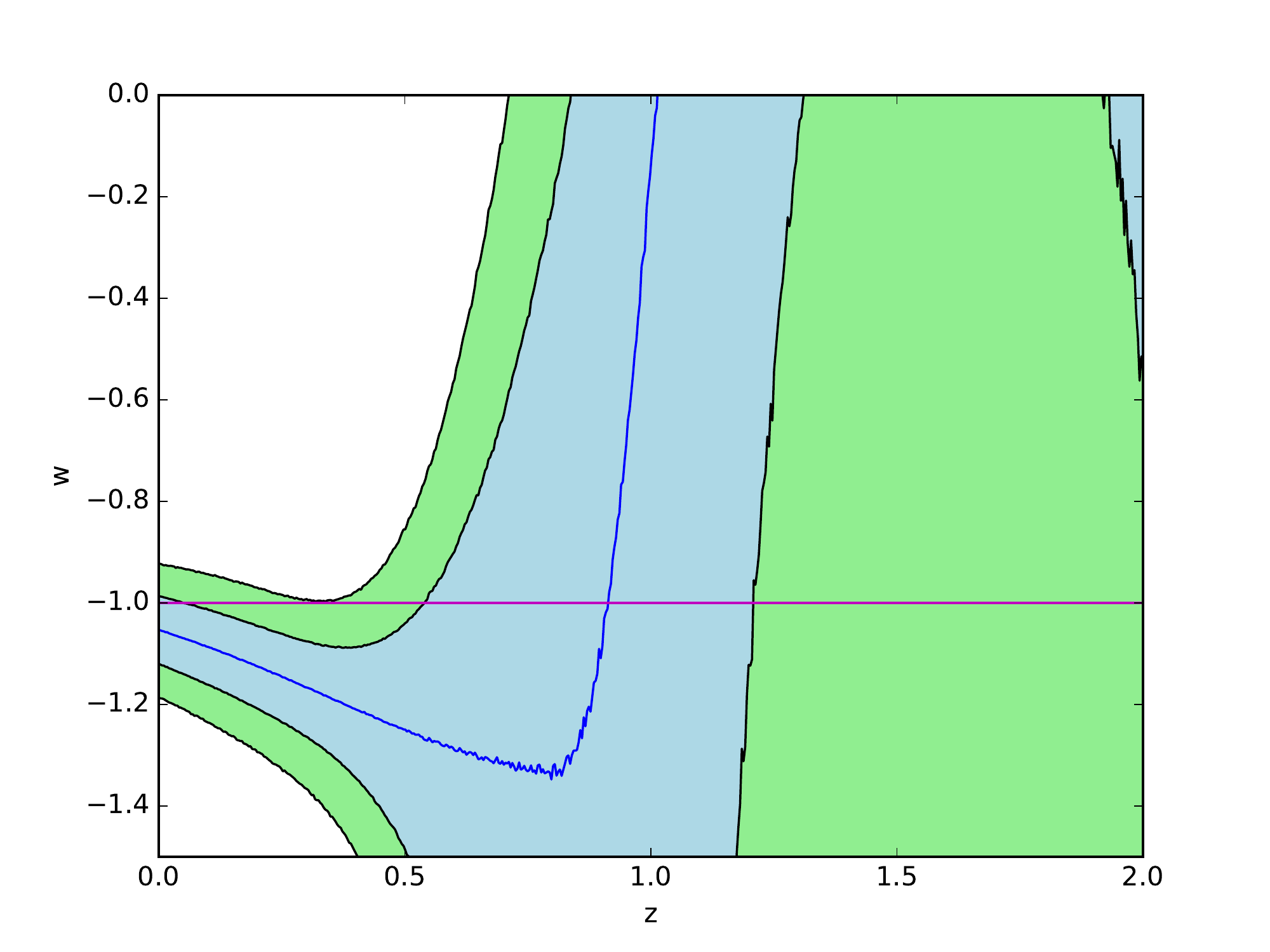}
\includegraphics[scale=0.23]{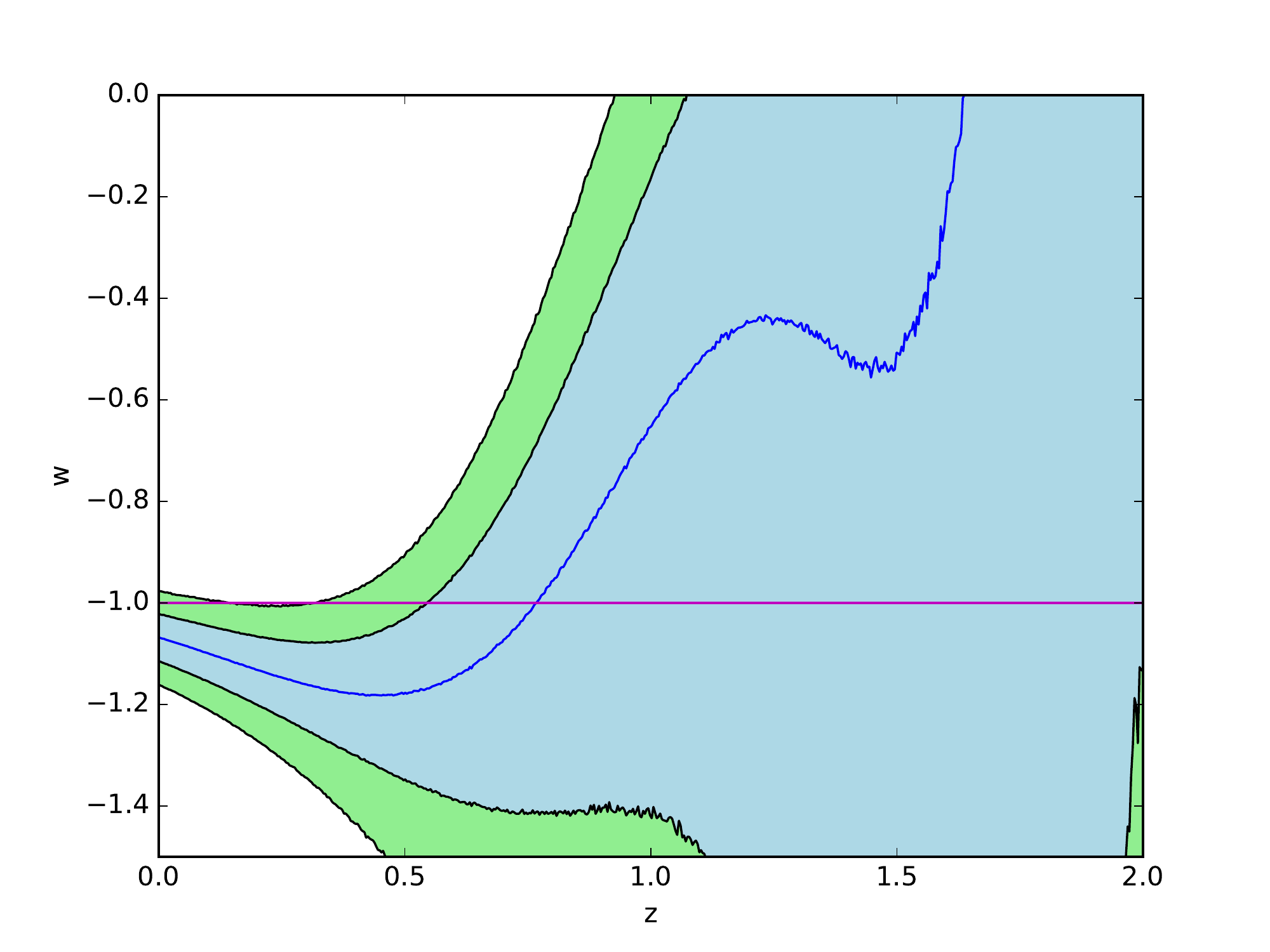}
\includegraphics[scale=0.23]{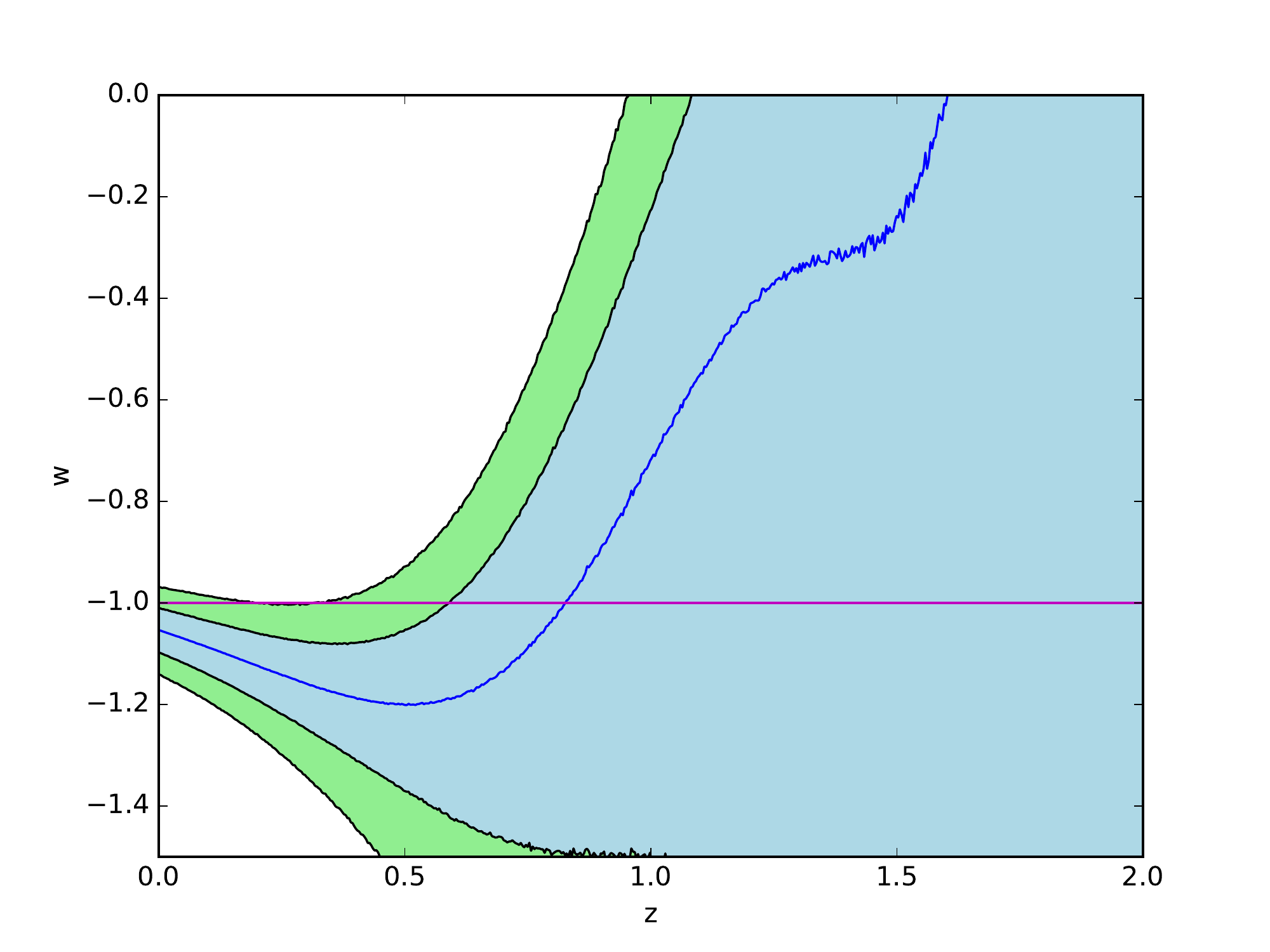}
\includegraphics[scale=0.23]{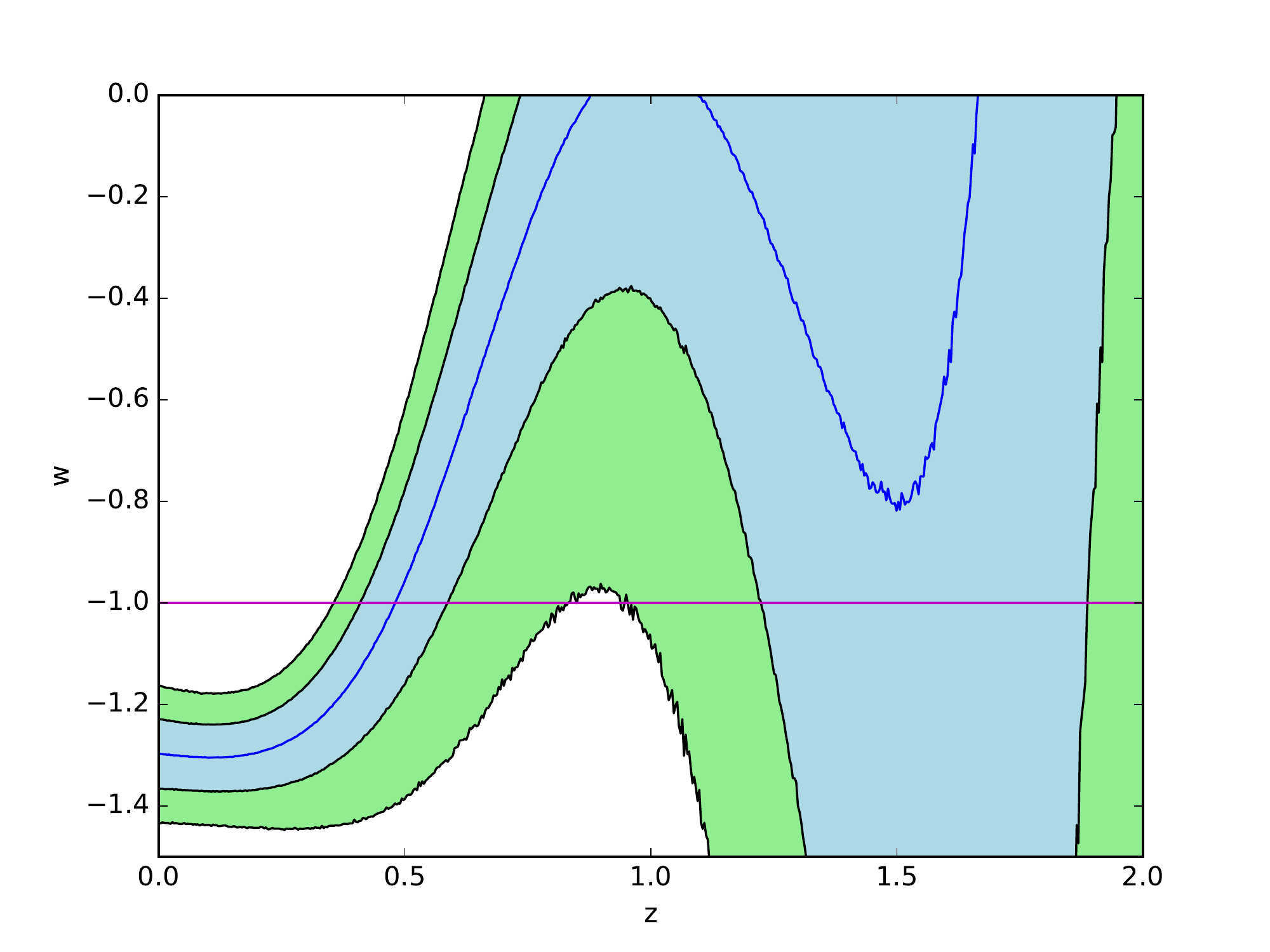}
\includegraphics[scale=0.23]{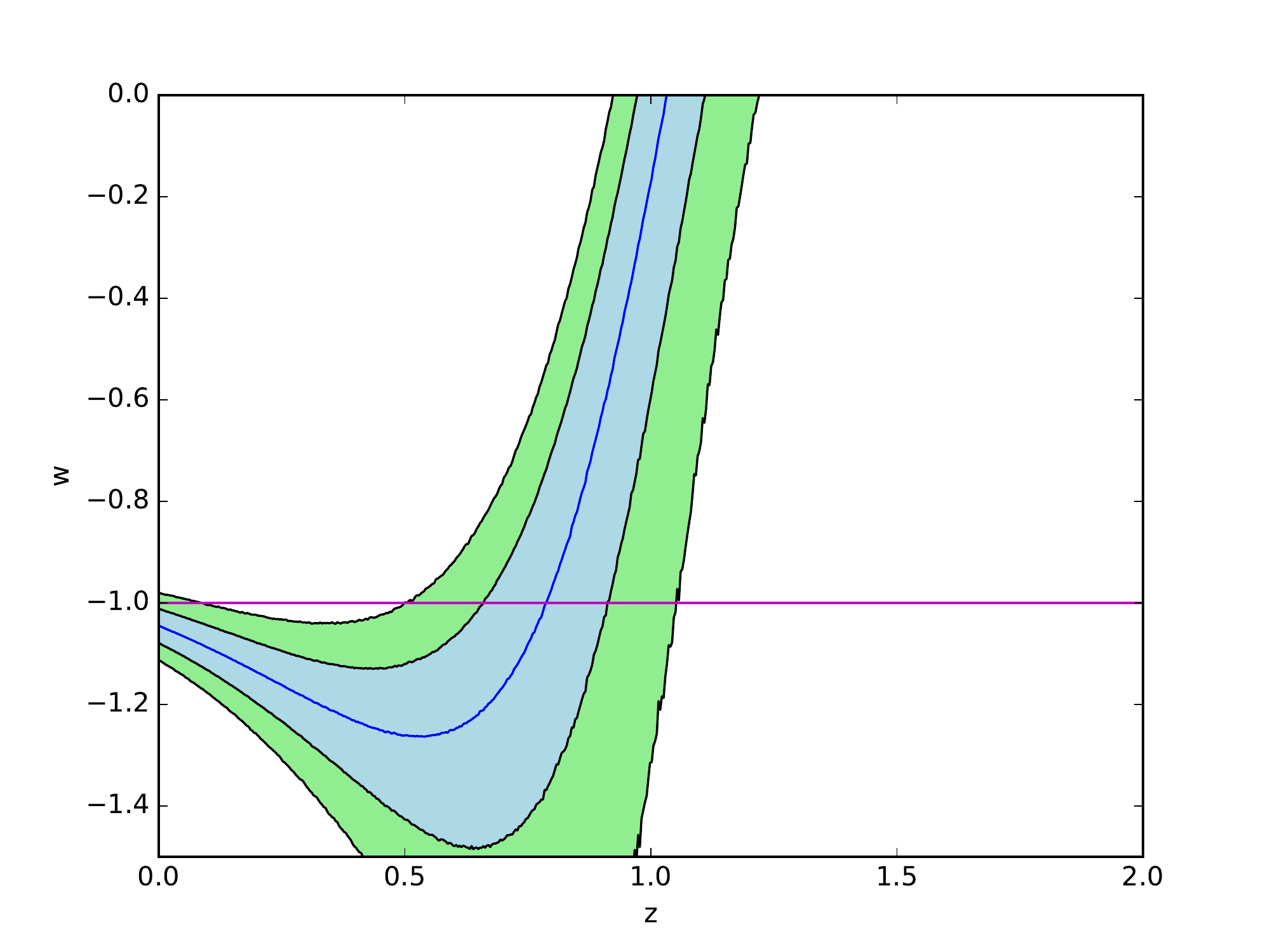}
\includegraphics[scale=0.23]{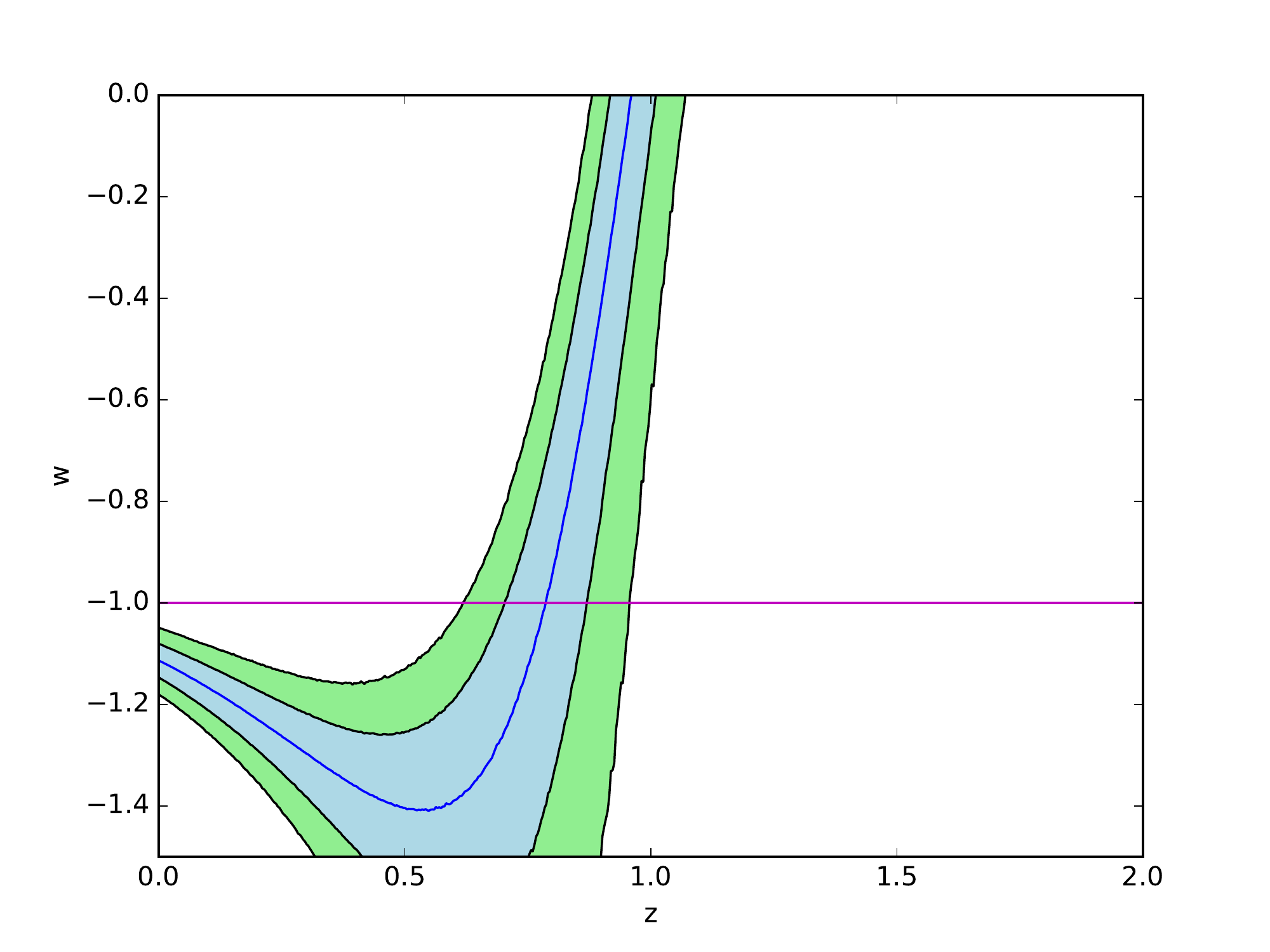}
\caption{The GP reconstructions of the EoS of DE $\omega(z)$ using different observations. The upper left panel, upper middle panel, upper right panel, lower left panel, lower middle panel and lower right panel correspond to JLA, JLA + H(z), JLA + H(z) + CMB, JLA + H(z) + CMB + HII, JLA + H(z) + CMB + GRB and JLA + H(z) + CMB + HII + GRB, respectively. The blue and magenta lines represents the underlying true model (the mean value of reconstructions) and the $\Lambda$CDM model, respectively. We have assumed $\Omega_{m0}=0.308\pm0.012$, $\Omega_{k0}=0$ and $73.24\pm1.74$ km s$^{-1}$ Mpc$^{-1}$.}\label{f3}
\end{figure}
\begin{figure}
\centering
\includegraphics[scale=0.4]{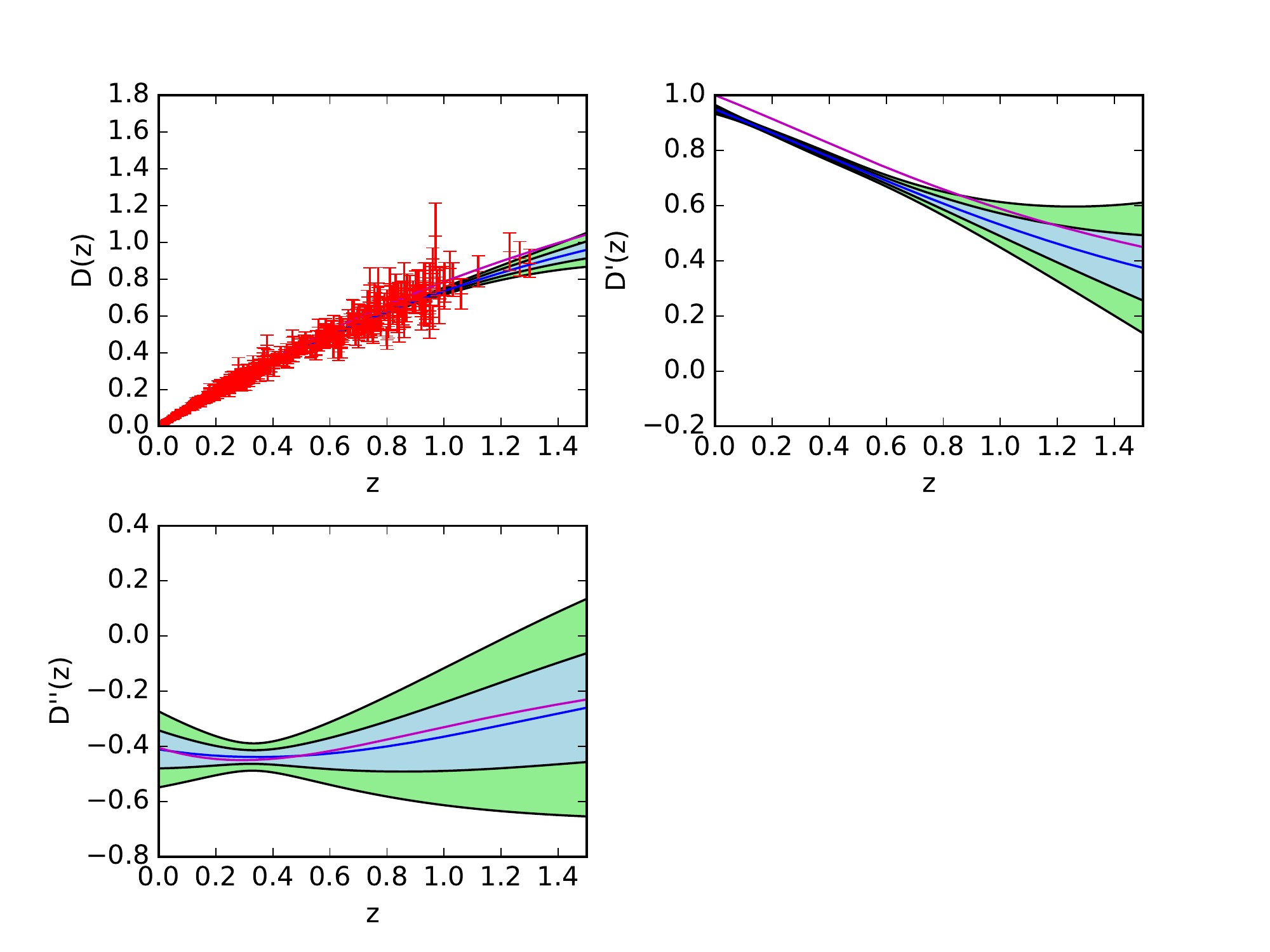}
\includegraphics[scale=0.4]{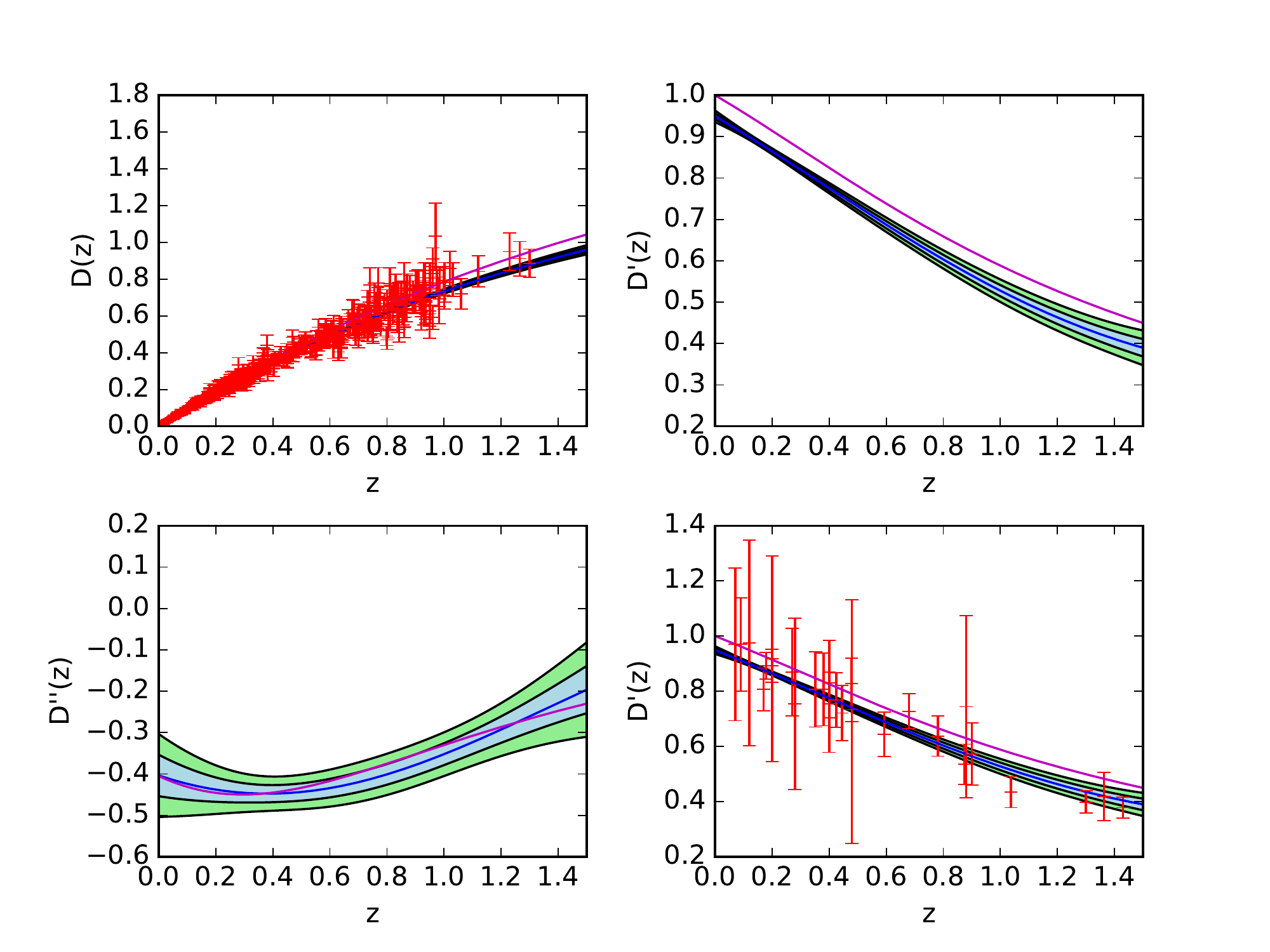}
\includegraphics[scale=0.4]{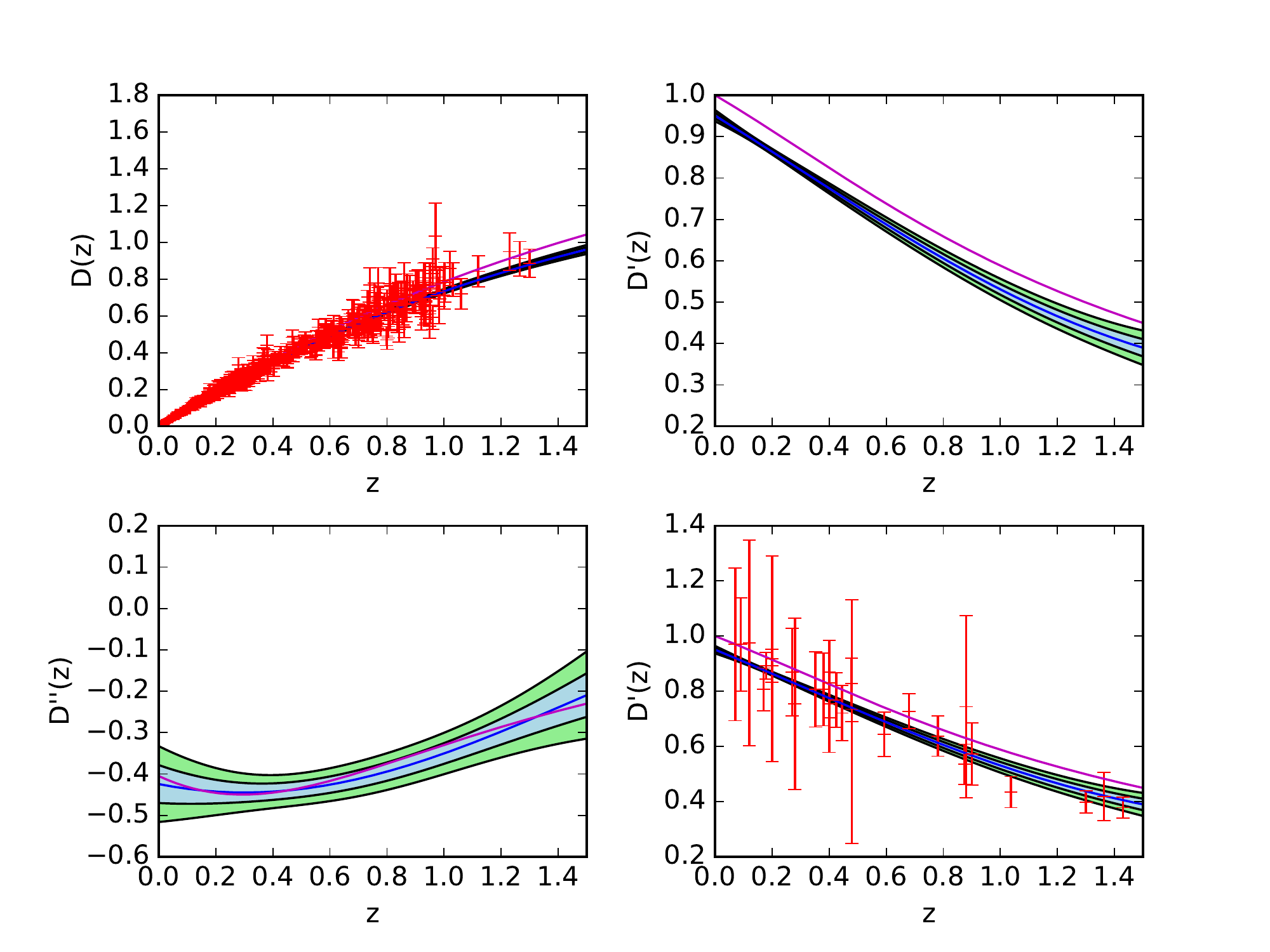}
\includegraphics[scale=0.4]{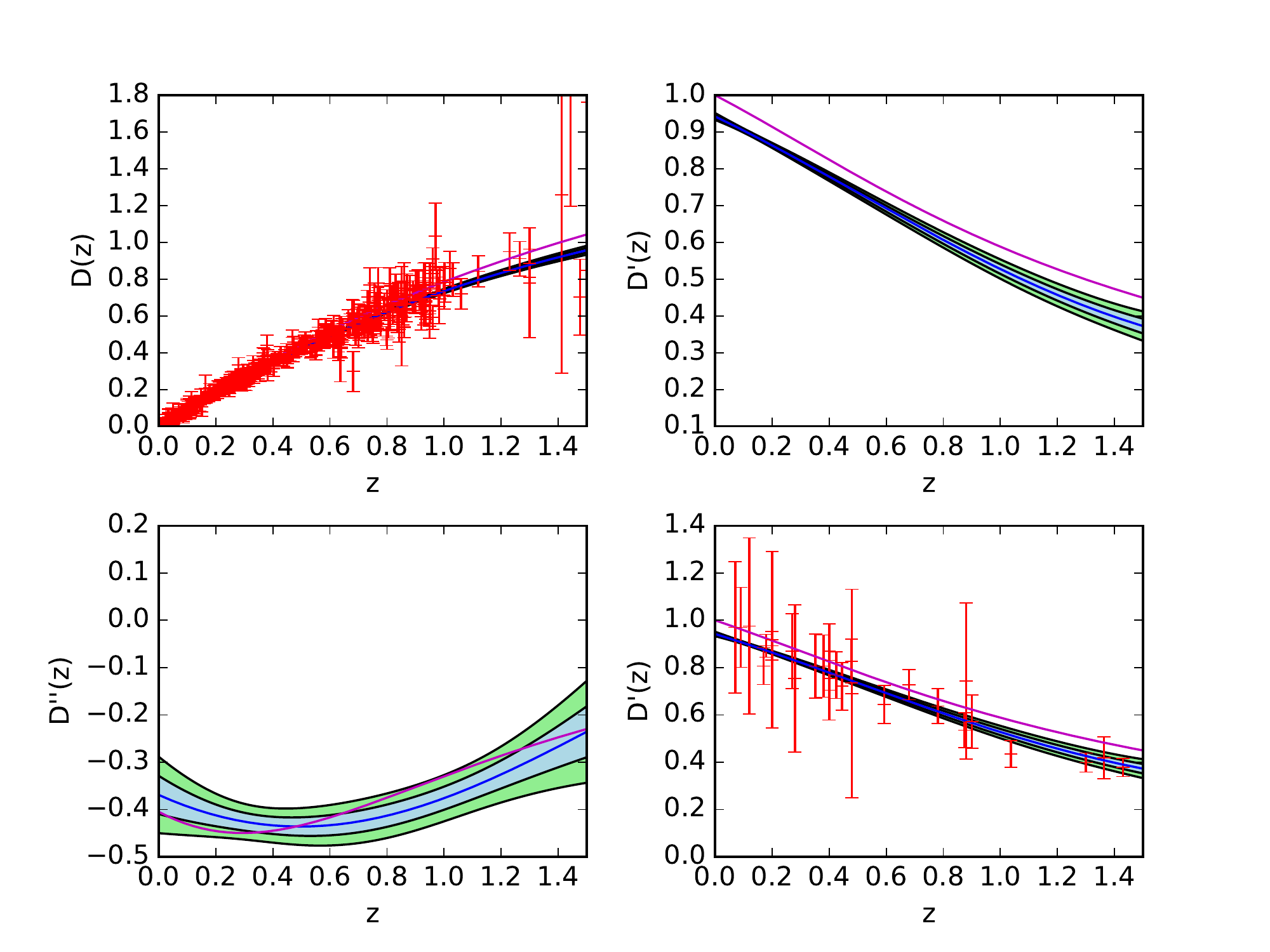}
\includegraphics[scale=0.4]{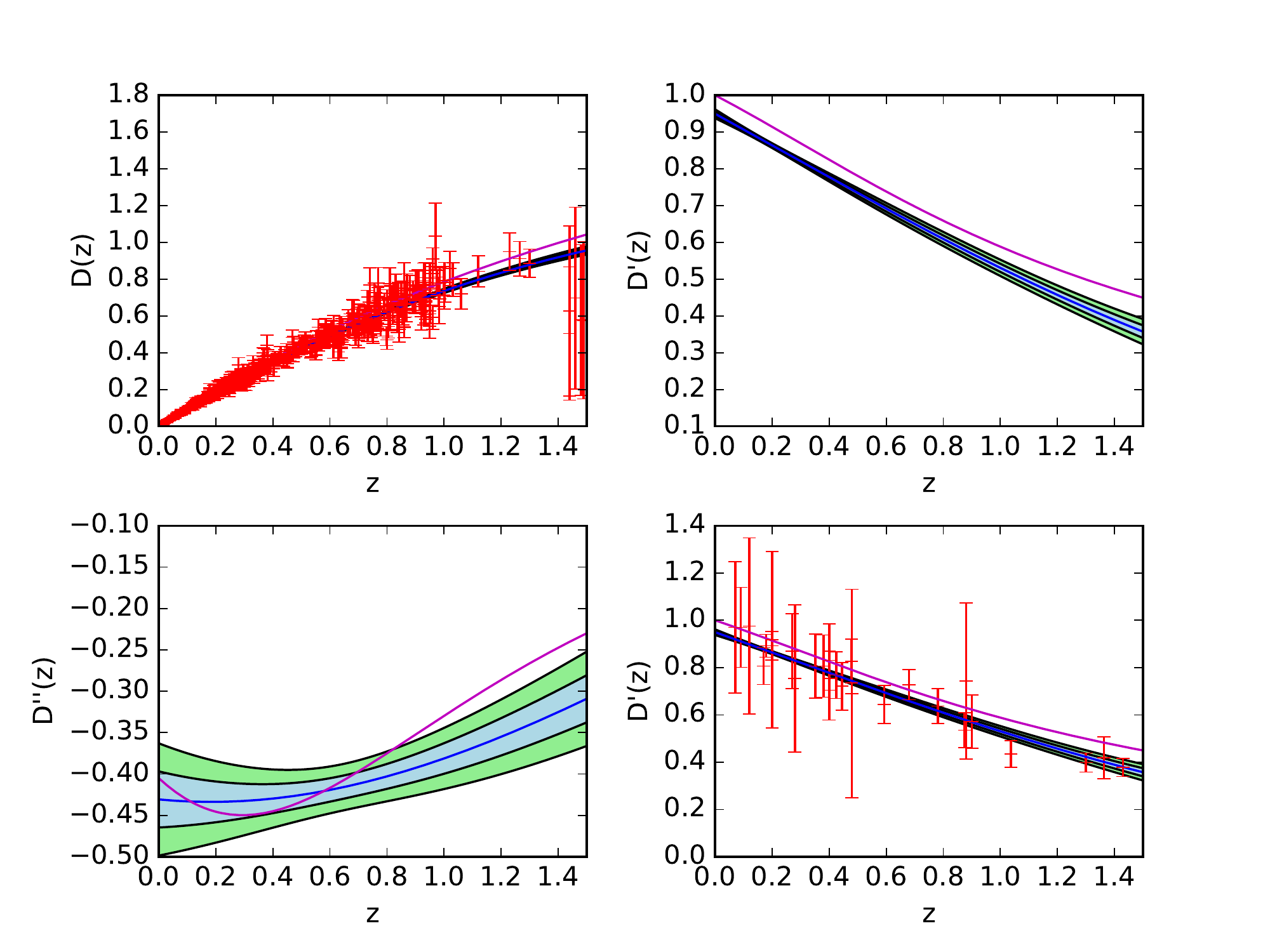}
\includegraphics[scale=0.4]{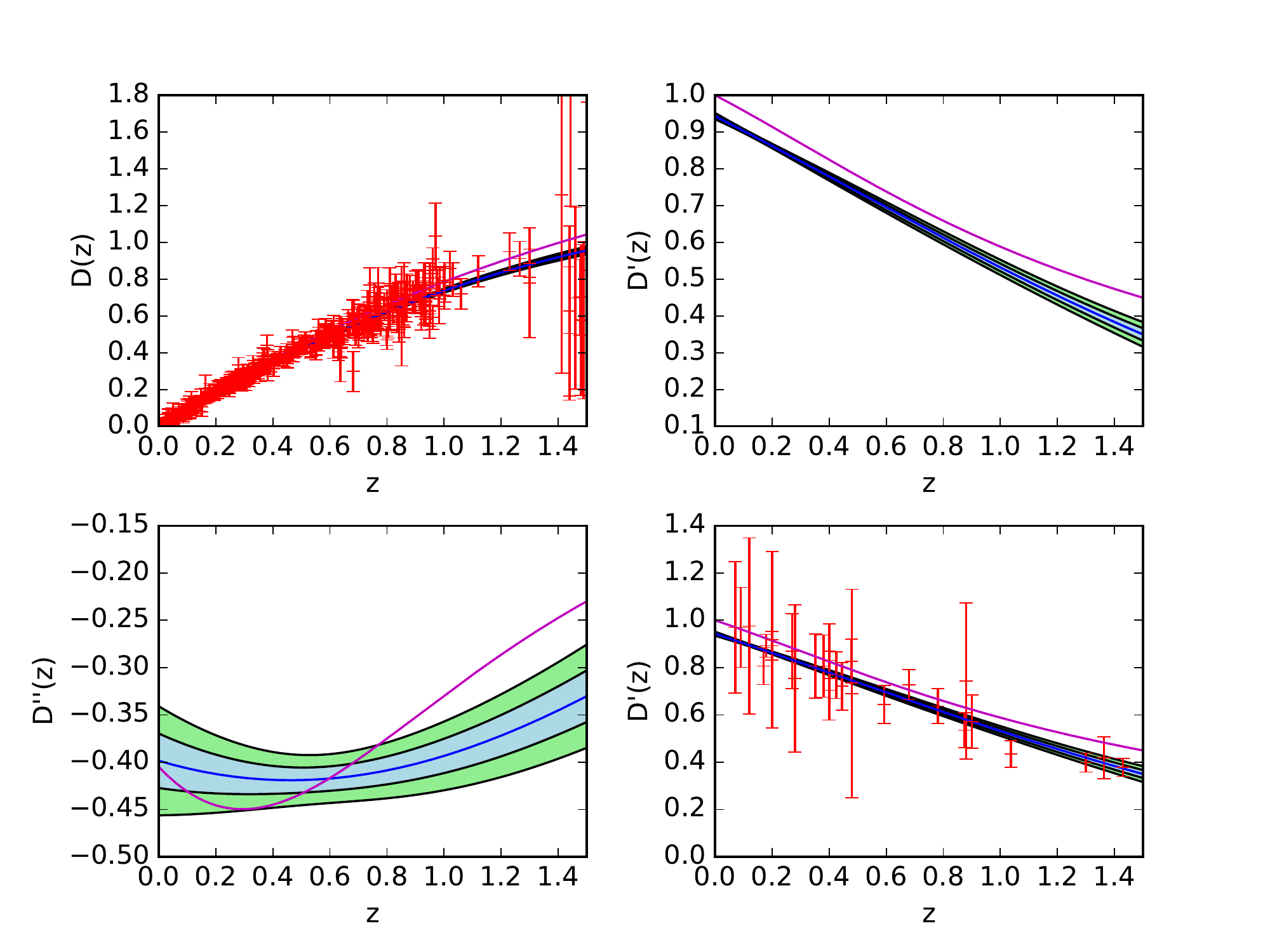}
\caption{The GP reconstructions of $D(z), D'(z)$ and $D''(z)$ using different observations. The upper left panel, upper right panel, middle left panel, middle right panel, lower left panel and lower right panel correspond to JLA, JLA + H(z), JLA + H(z) + CMB, JLA + H(z) + CMB + HII, JLA + H(z) + CMB + GRB and JLA + H(z) + CMB + HII + GRB, respectively. We have assumed $\Omega_{m0}=0.308\pm0.012$, $\Omega_{k0}=0$ and $66.93\pm0.62$ km s$^{-1}$ Mpc$^{-1}$.}\label{f4}
\end{figure}
\begin{figure}
\centering
\includegraphics[scale=0.23]{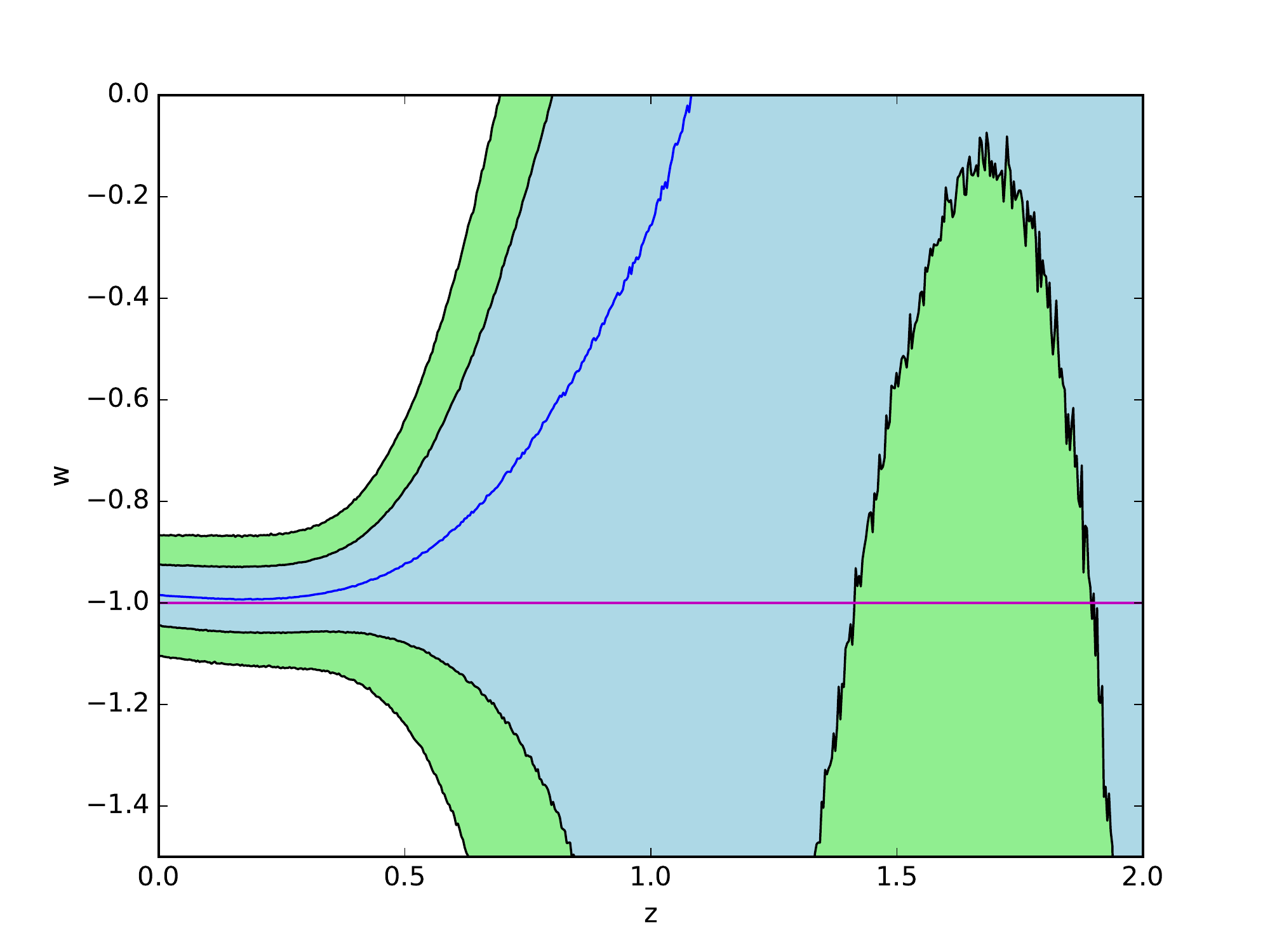}
\includegraphics[scale=0.23]{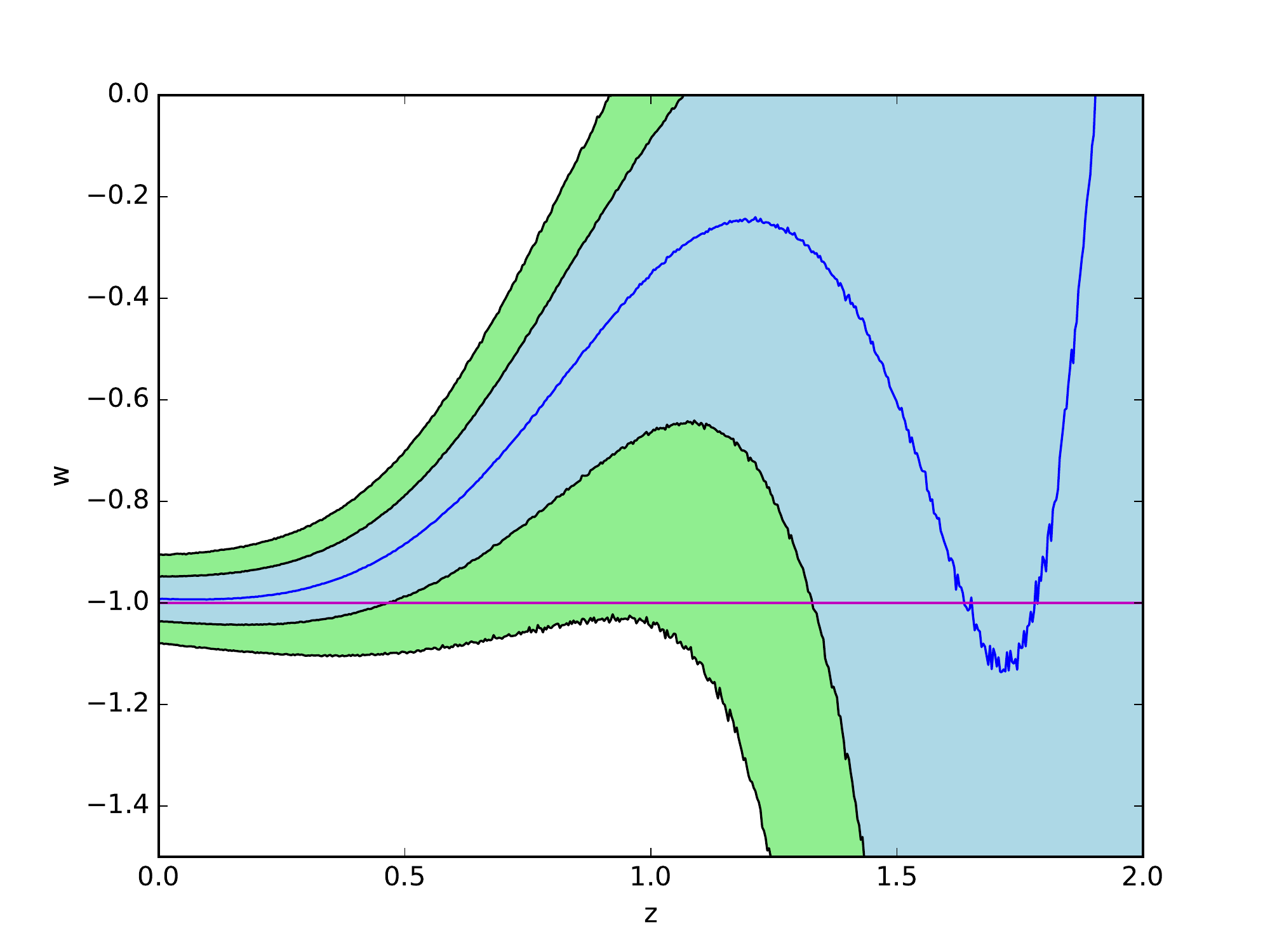}
\includegraphics[scale=0.23]{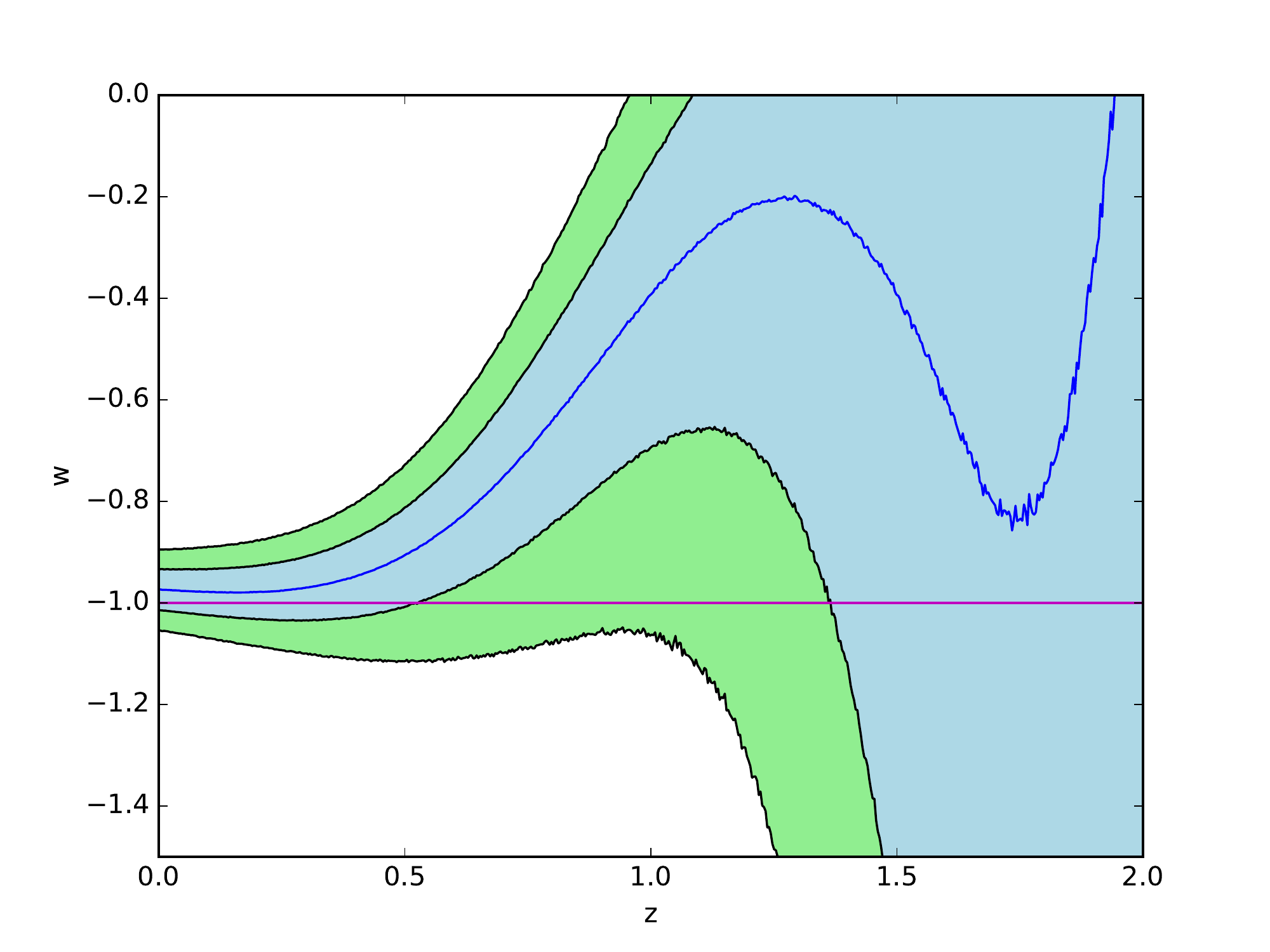}
\includegraphics[scale=0.23]{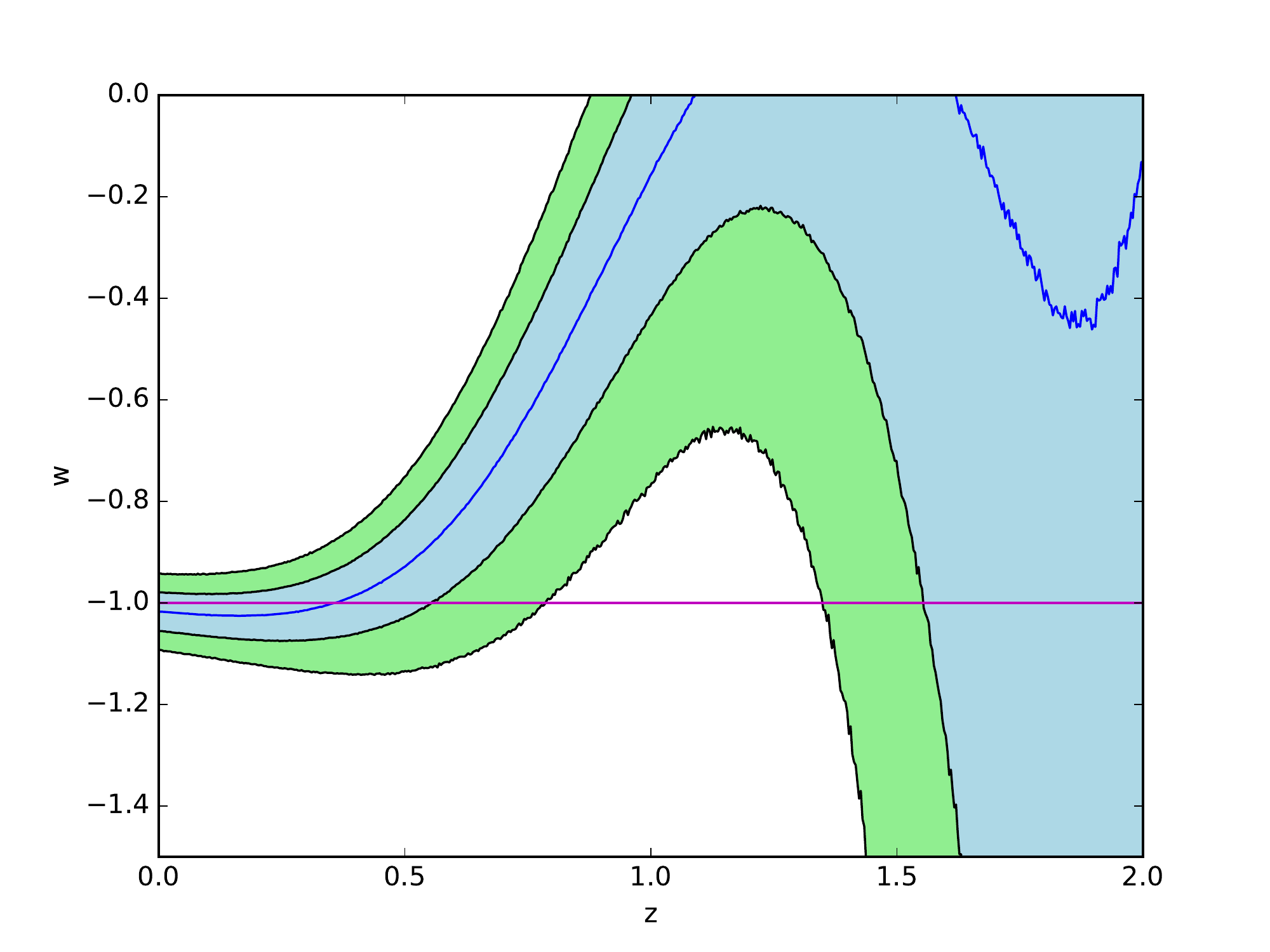}
\includegraphics[scale=0.23]{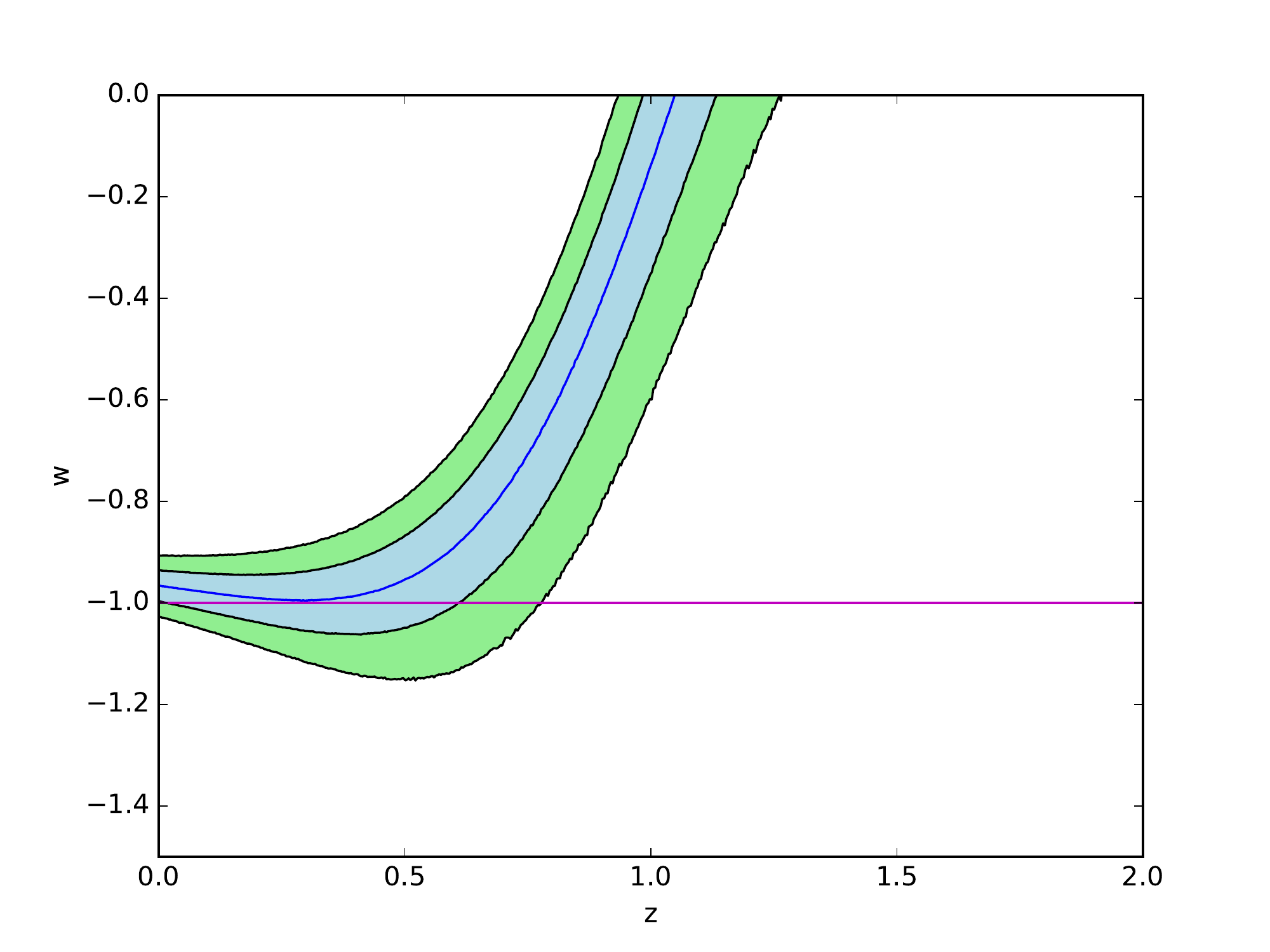}
\includegraphics[scale=0.23]{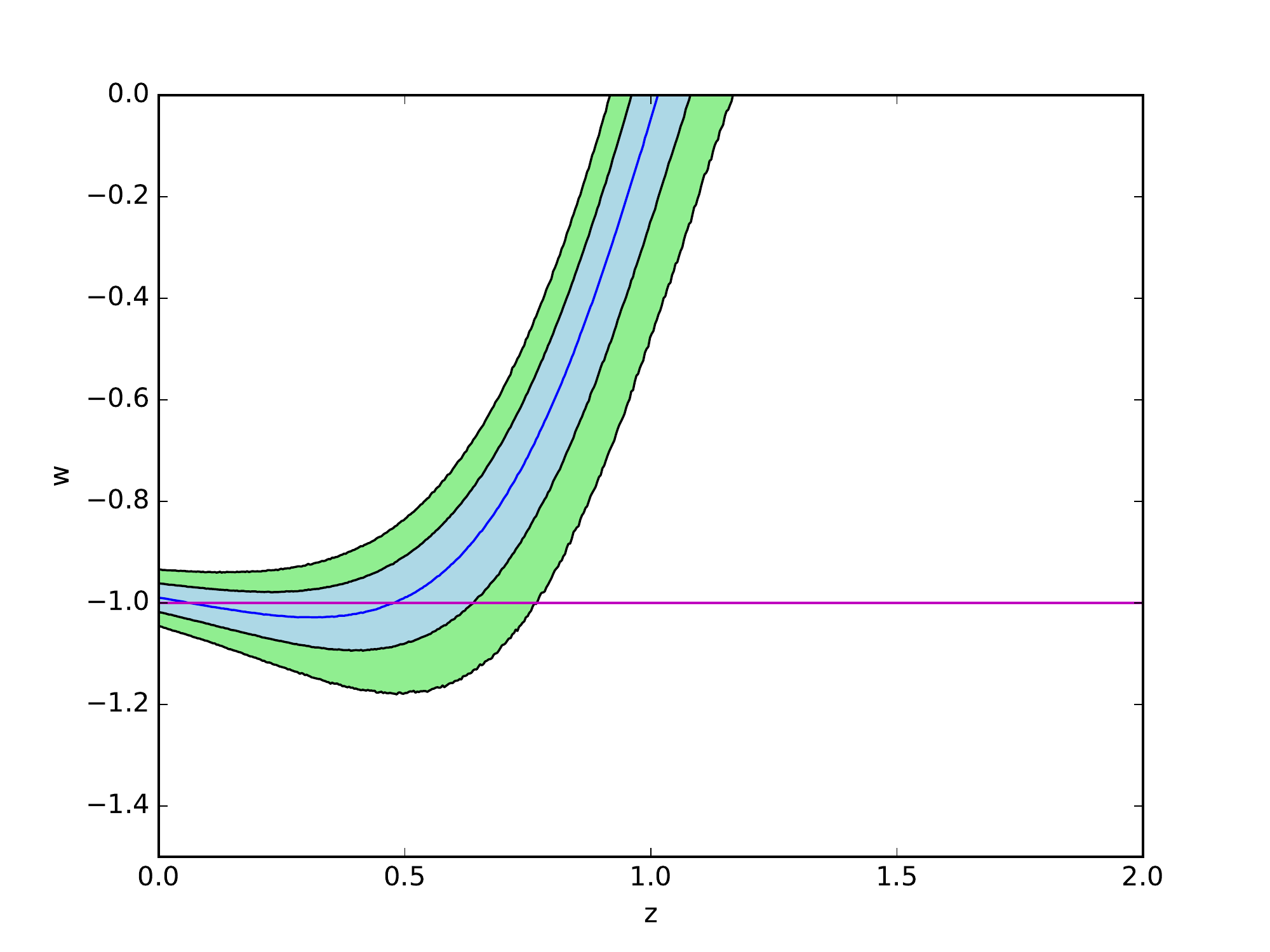}
\caption{The GP reconstructions of the EoS of DE $\omega(z)$ using different observations. The upper left panel, upper middle panel, upper right panel, lower left panel, lower middle panel and lower right panel correspond to JLA, JLA + H(z), JLA + H(z) + CMB, JLA + H(z) + CMB + HII, JLA + H(z) + CMB + GRB and JLA + H(z) + CMB + HII + GRB, respectively. We have assumed $\Omega_{m0}=0.308\pm0.012$, $\Omega_{k0}=0$ and $66.93\pm0.62$ km s$^{-1}$ Mpc$^{-1}$.}\label{f5}
\end{figure}
\begin{figure}
\centering
\includegraphics[scale=0.23]{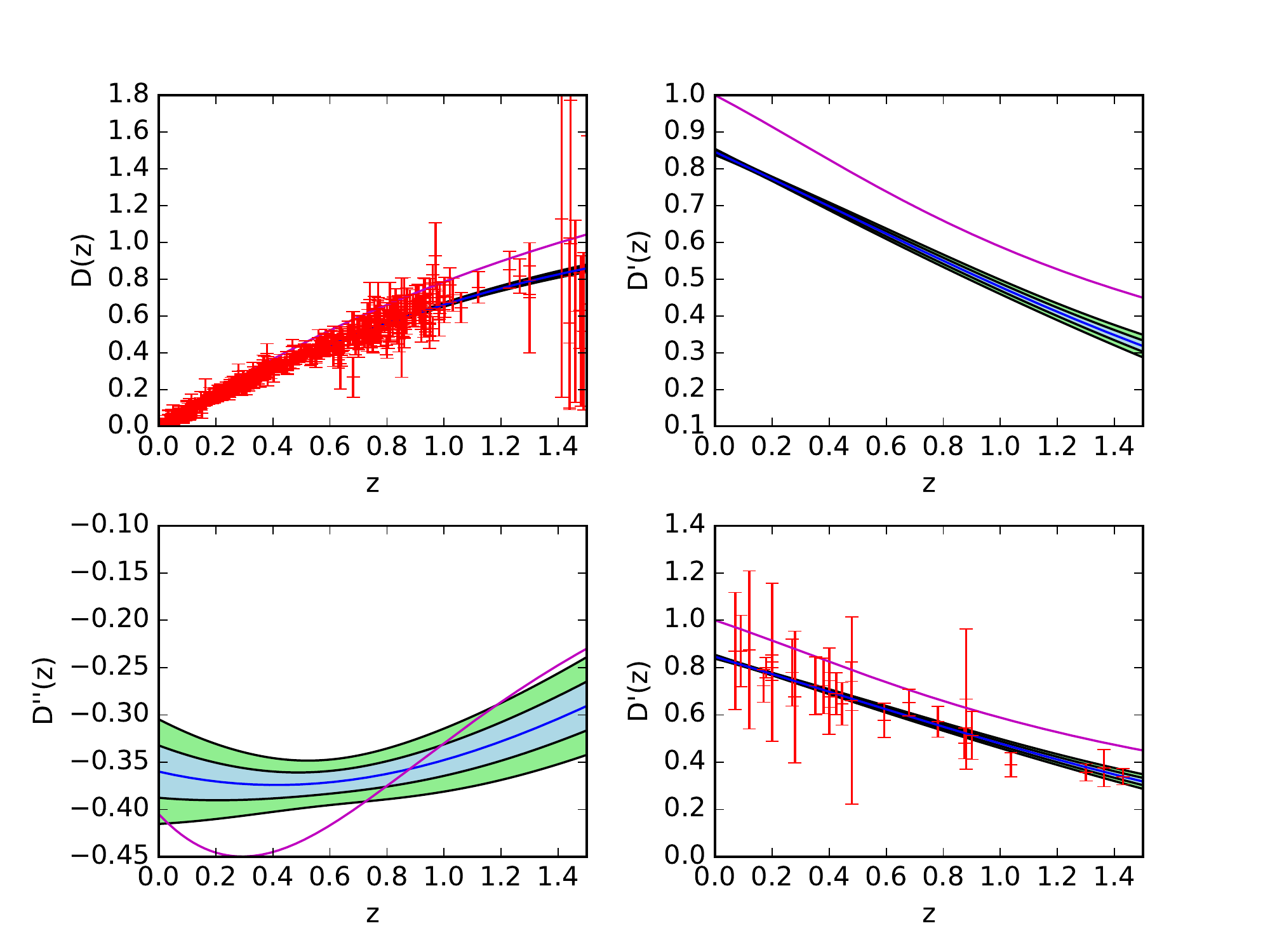}
\includegraphics[scale=0.23]{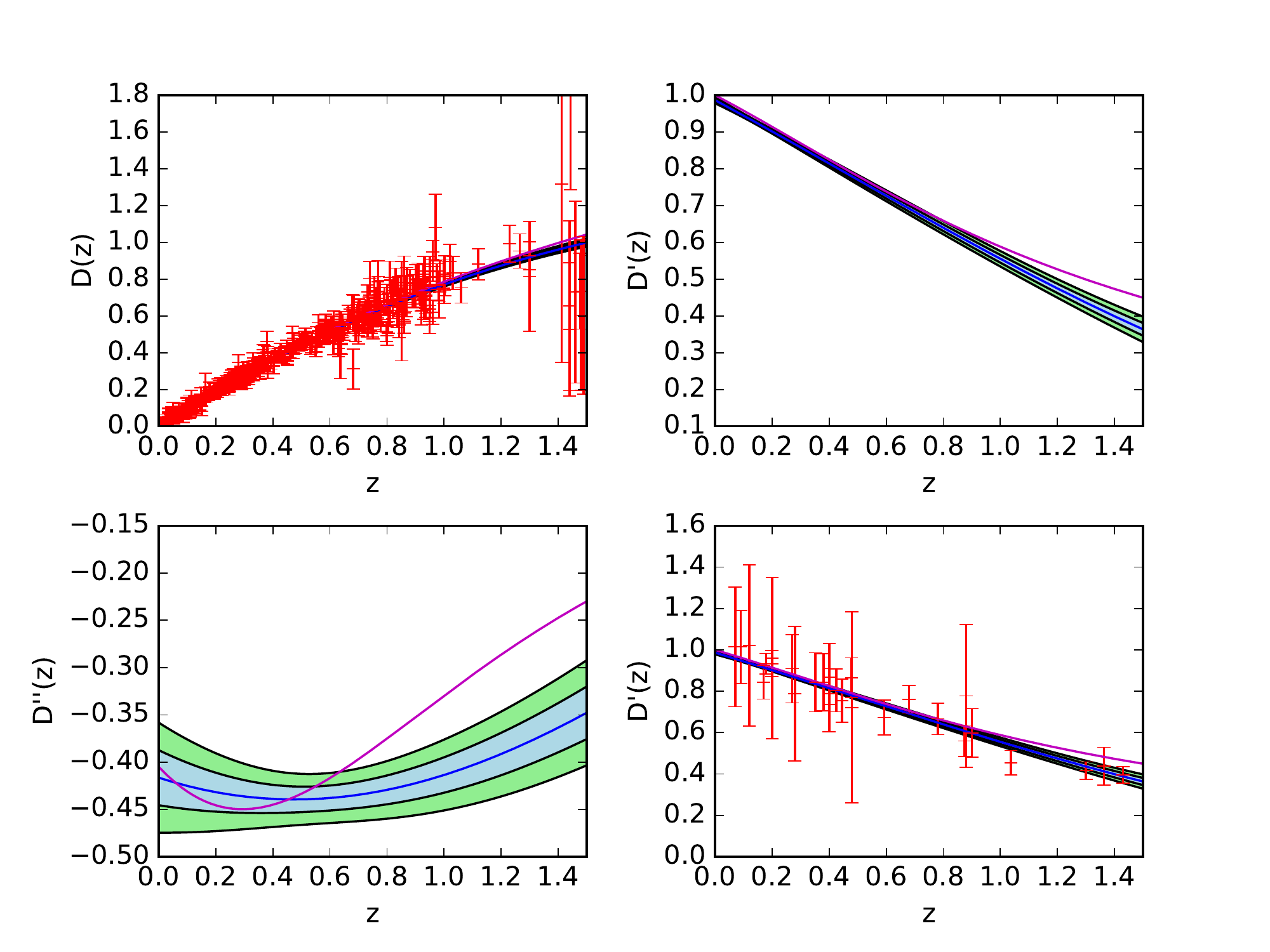}
\includegraphics[scale=0.23]{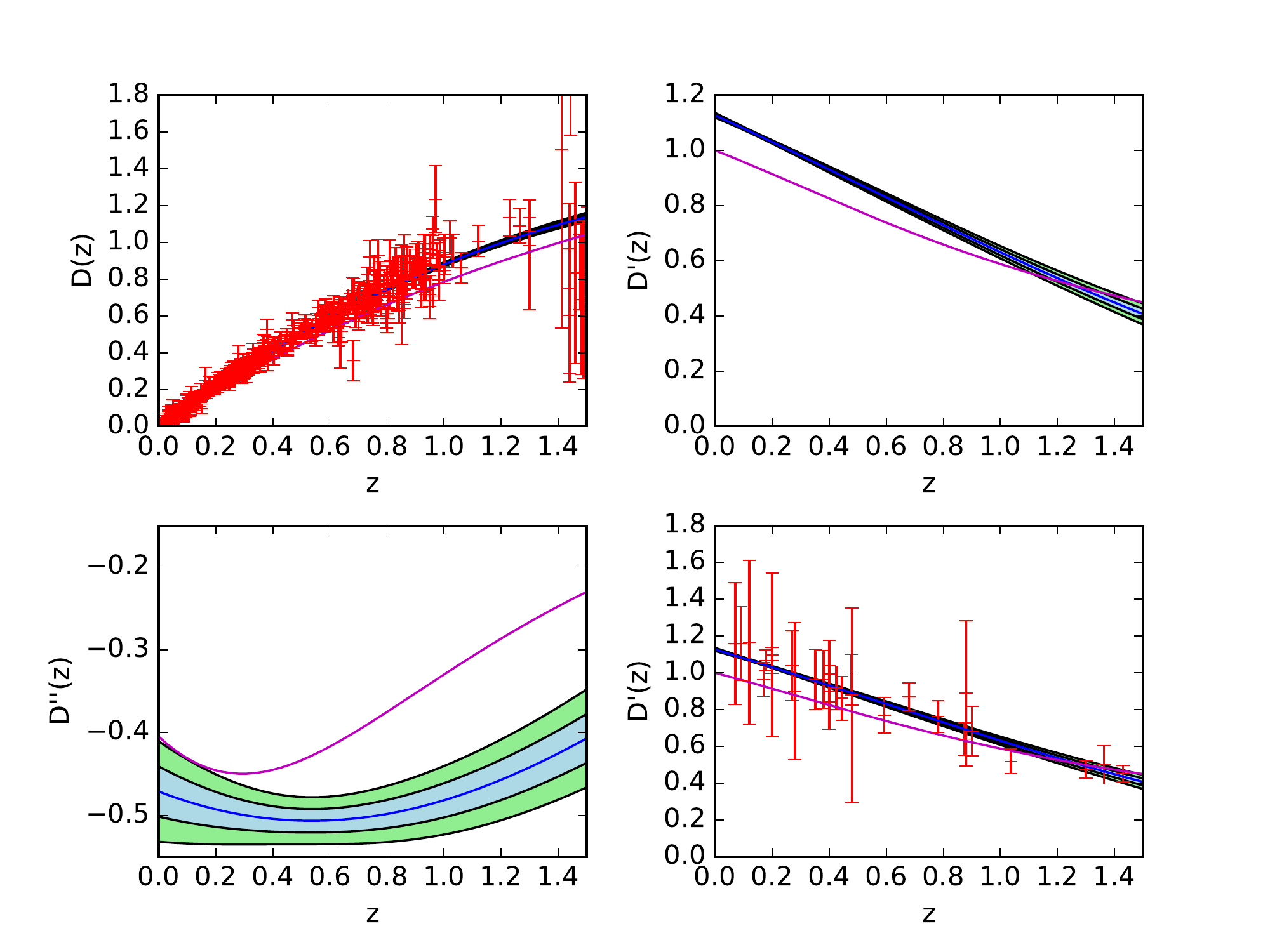}
\includegraphics[scale=0.23]{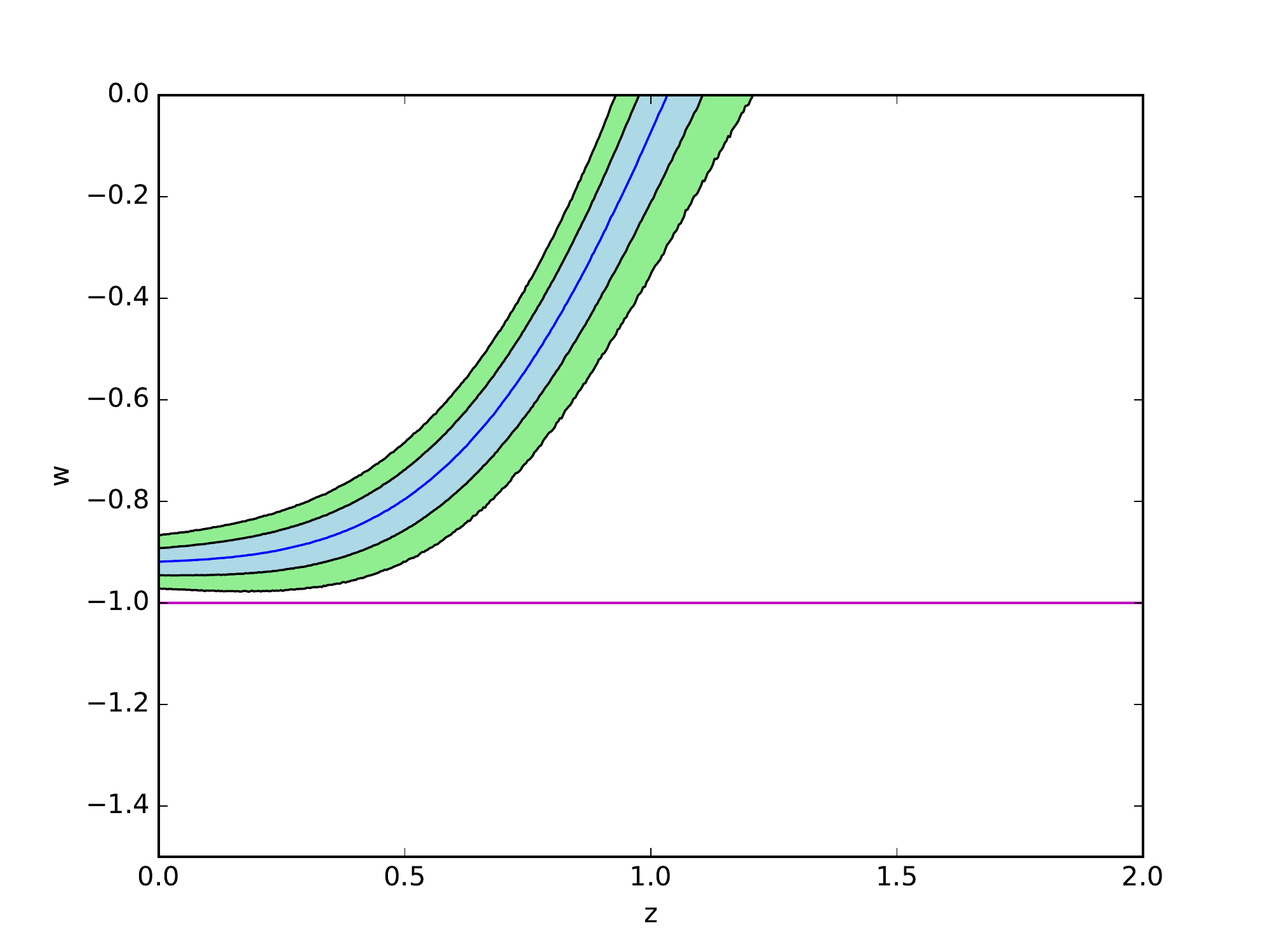}
\includegraphics[scale=0.23]{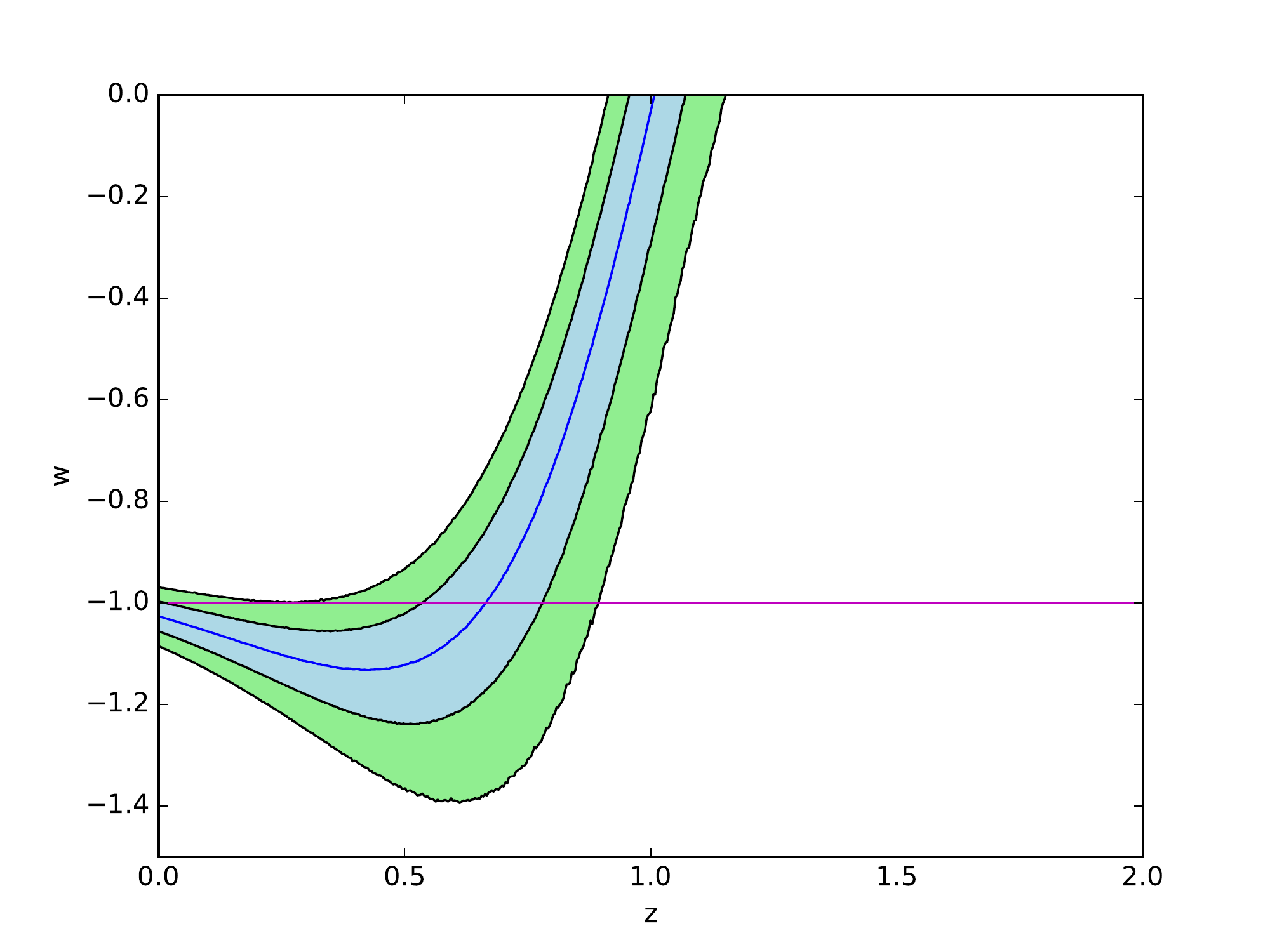}
\includegraphics[scale=0.23]{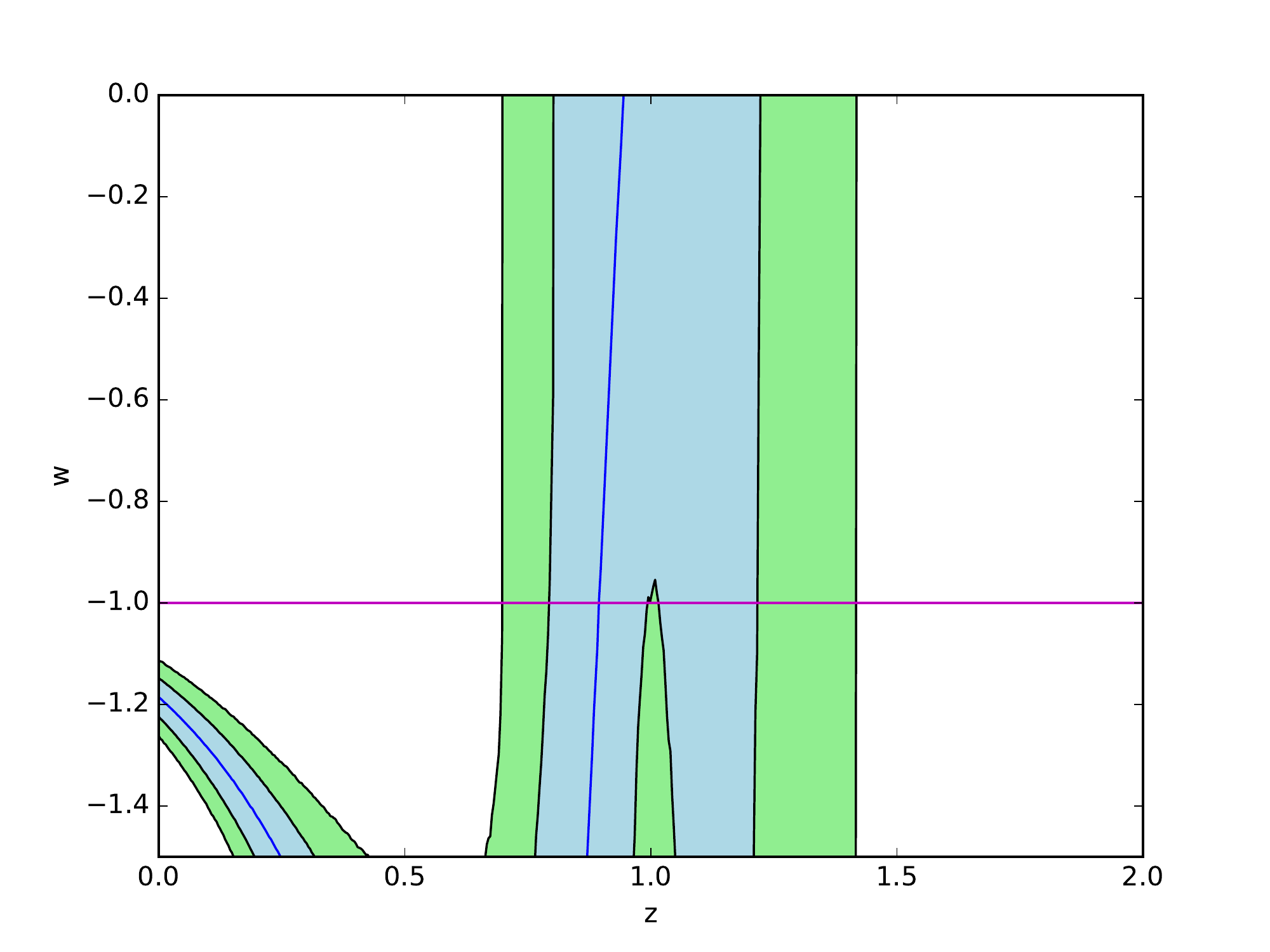}
\caption{The GP reconstructions of $D(z), D'(z)$, $D''(z)$ and the EoS of DE using JLA + H(z) + CMB + HII + GRB. The upper and lower panels from left to right correspond to the cases of $H_0=60$, $70$ and $80$ km s$^{-1}$ Mpc$^{-1}$. We have assumed $\Omega_{m0}=0.308\pm0.012$ and $\Omega_{k0}=0$.}\label{f6}
\end{figure}
\begin{figure}
\centering
\includegraphics[scale=0.23]{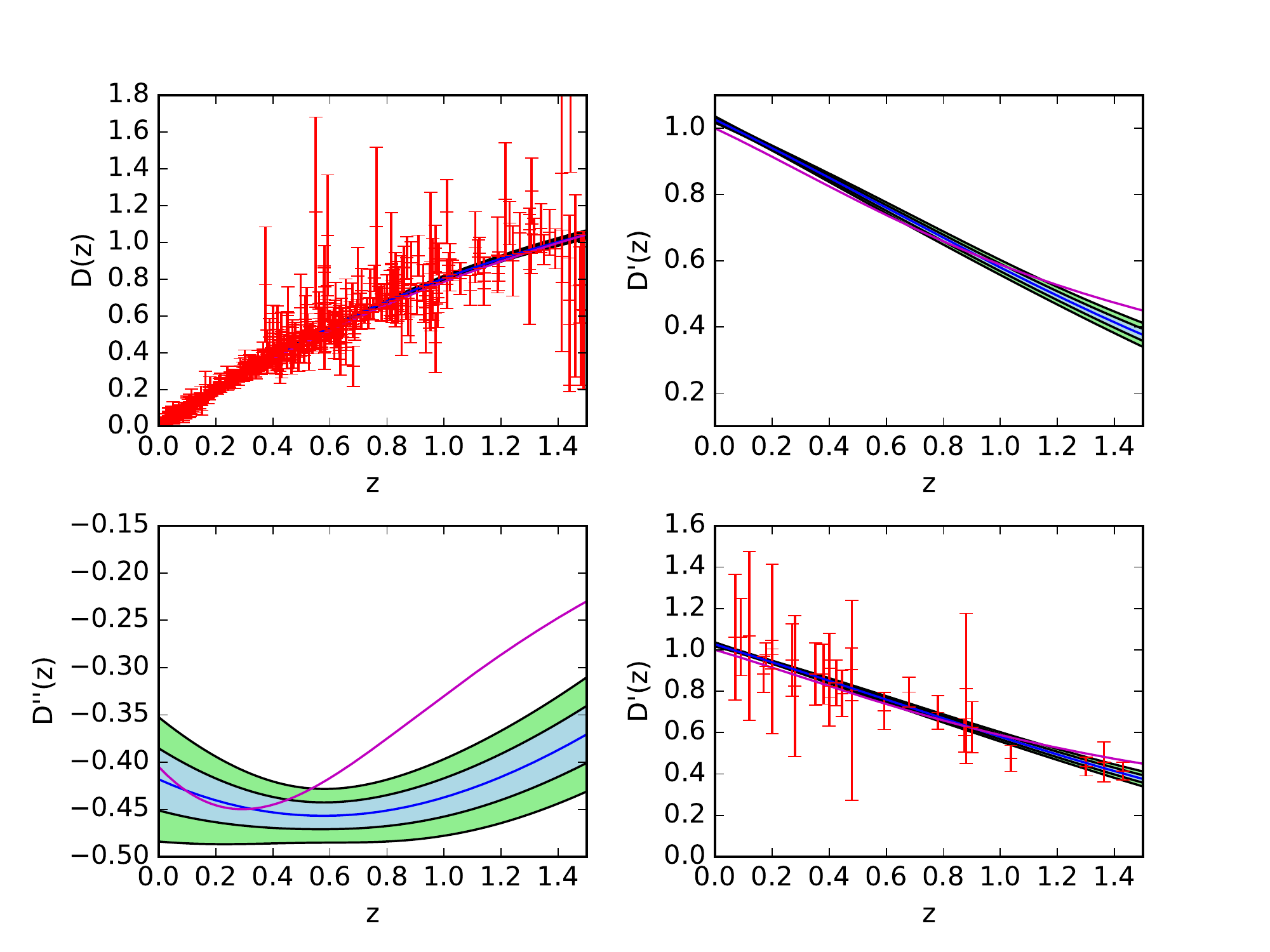}
\includegraphics[scale=0.23]{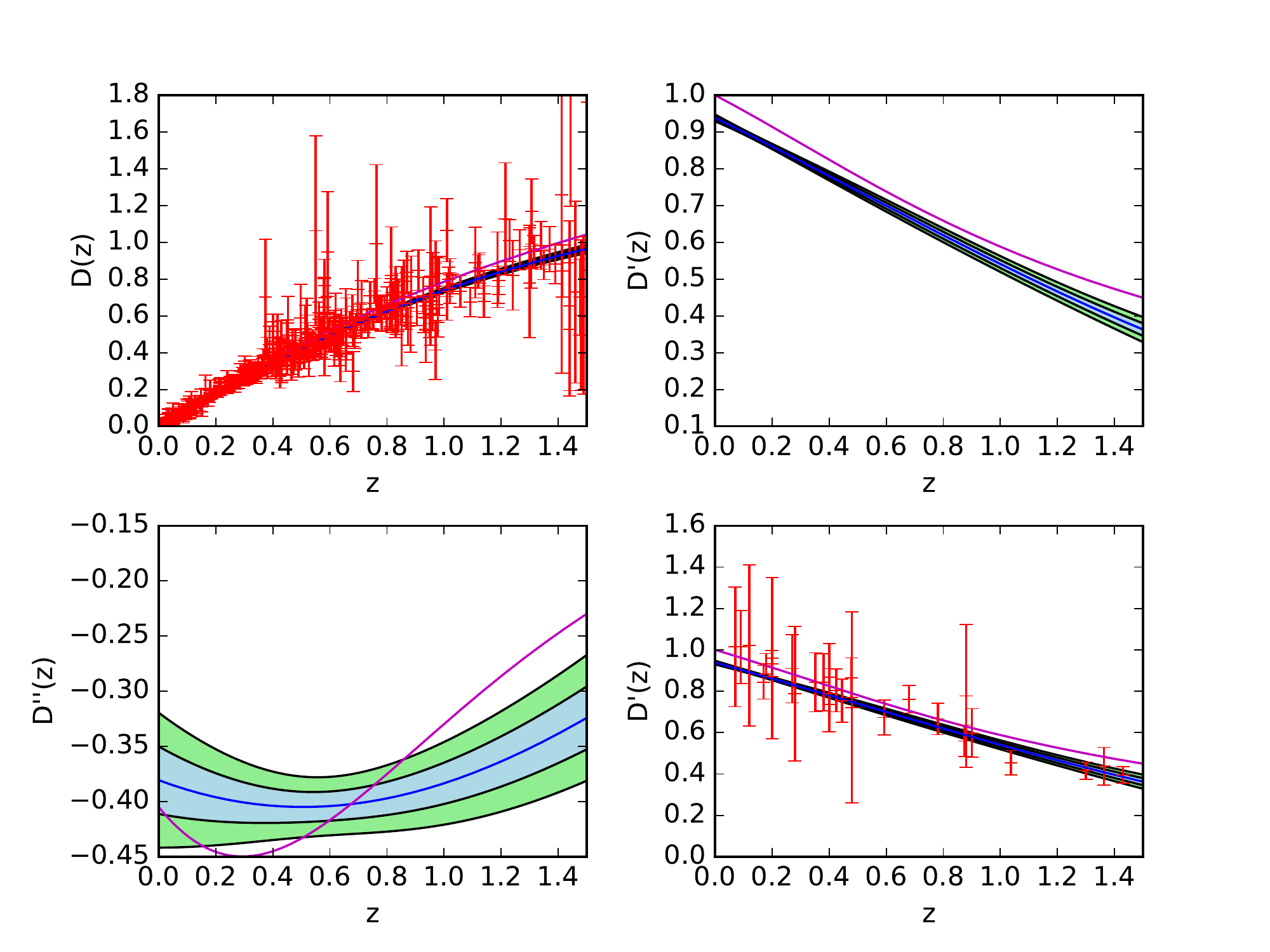}
\includegraphics[scale=0.23]{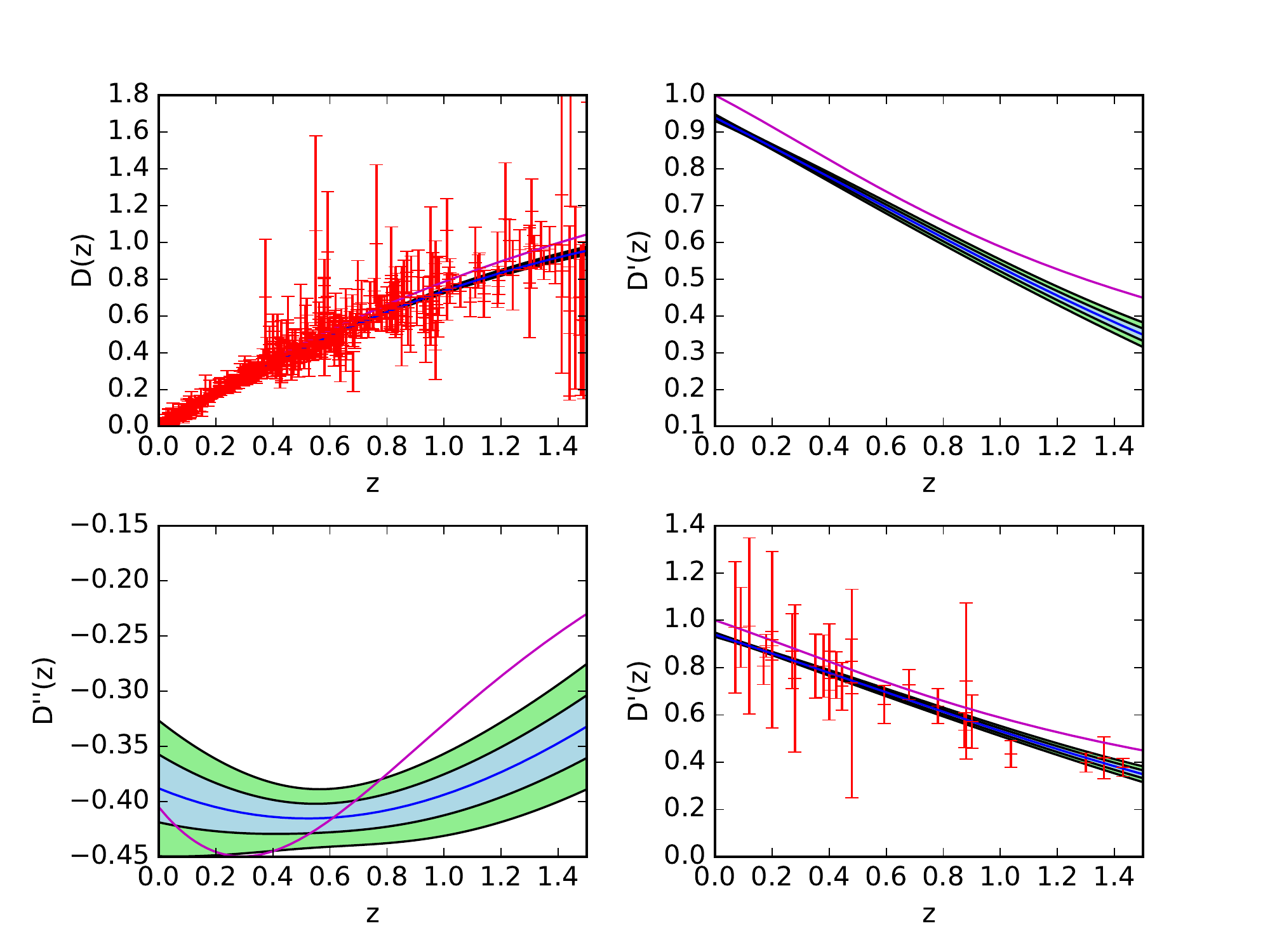}
\includegraphics[scale=0.23]{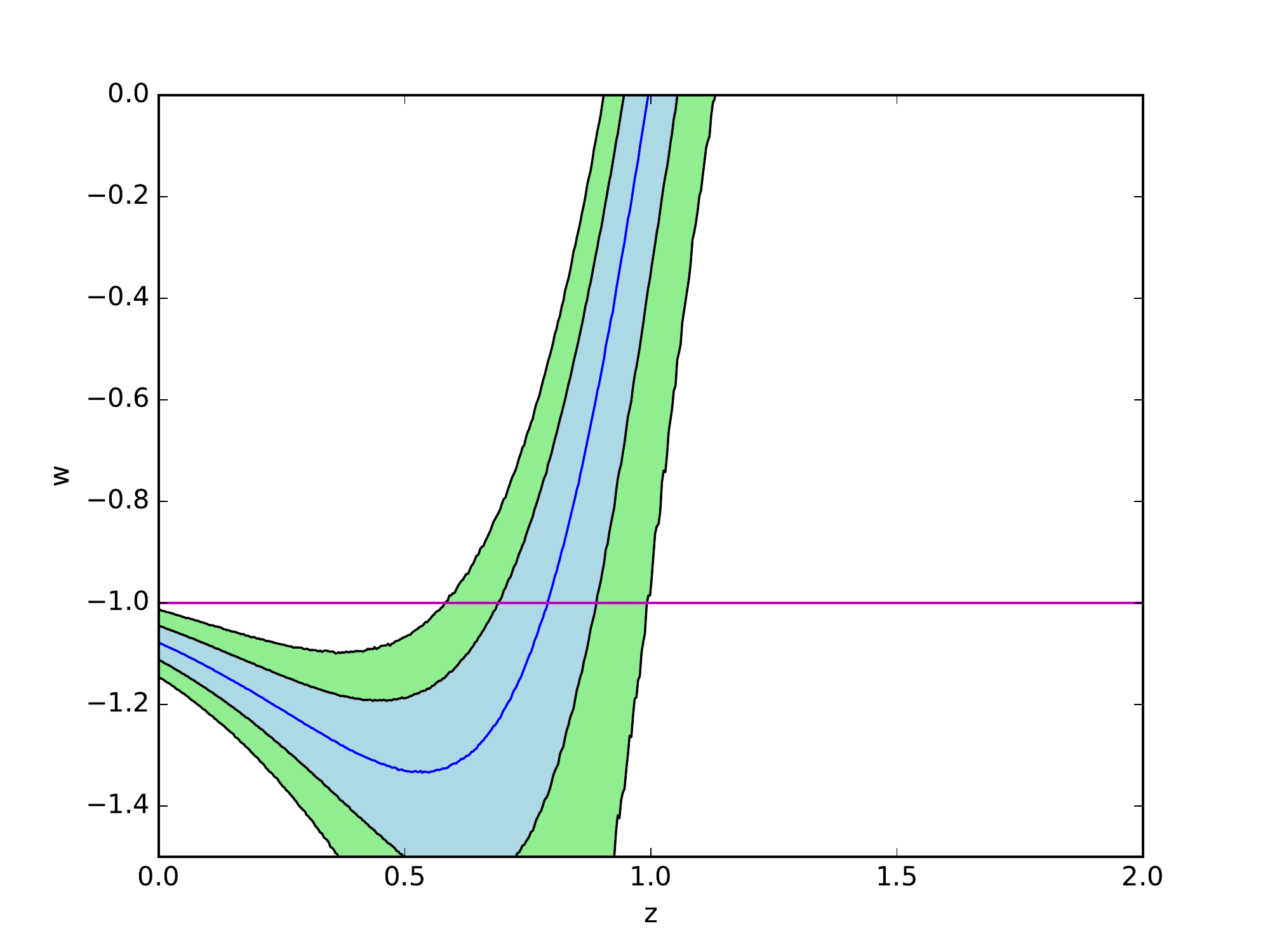}
\includegraphics[scale=0.23]{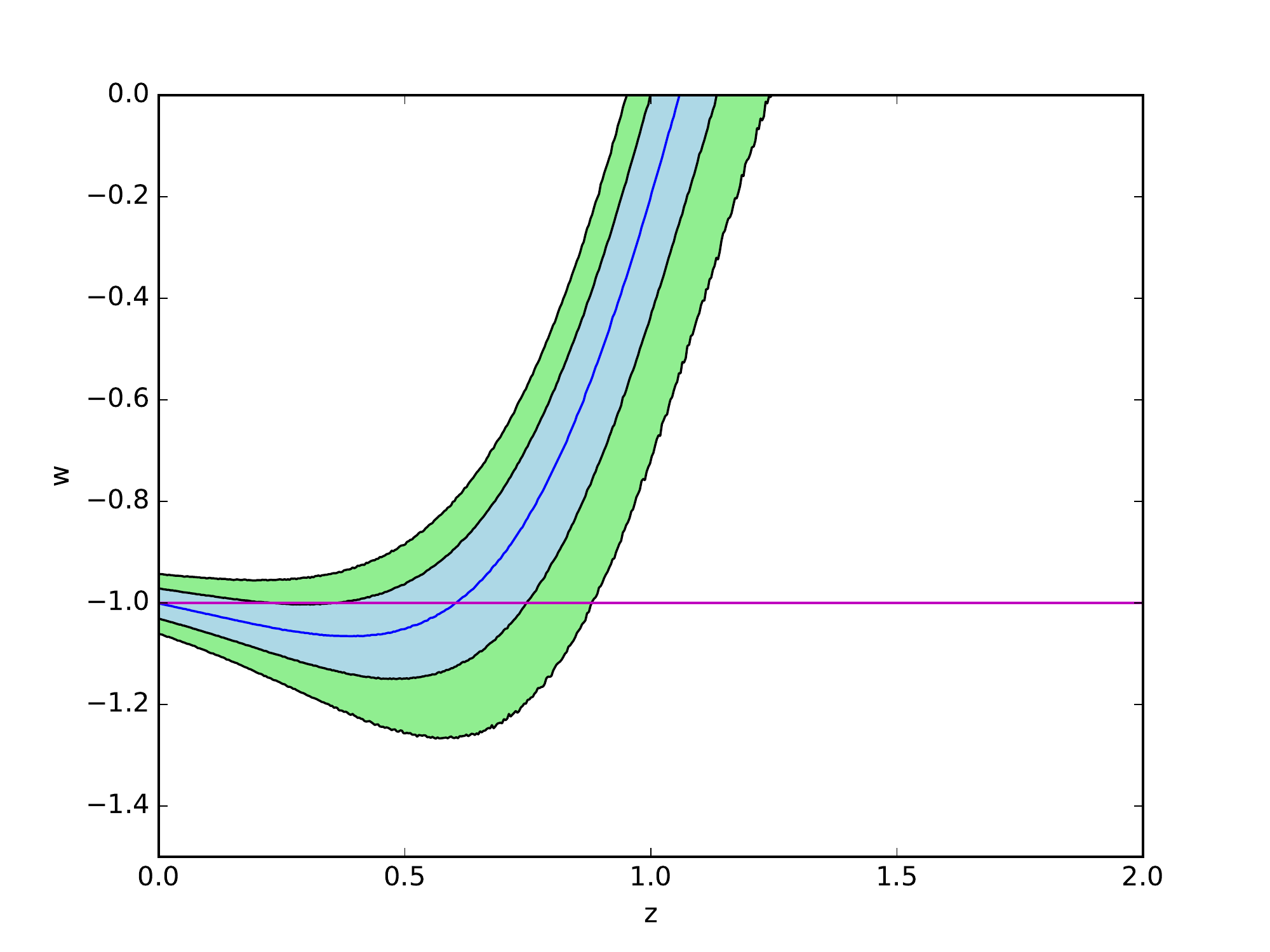}
\includegraphics[scale=0.23]{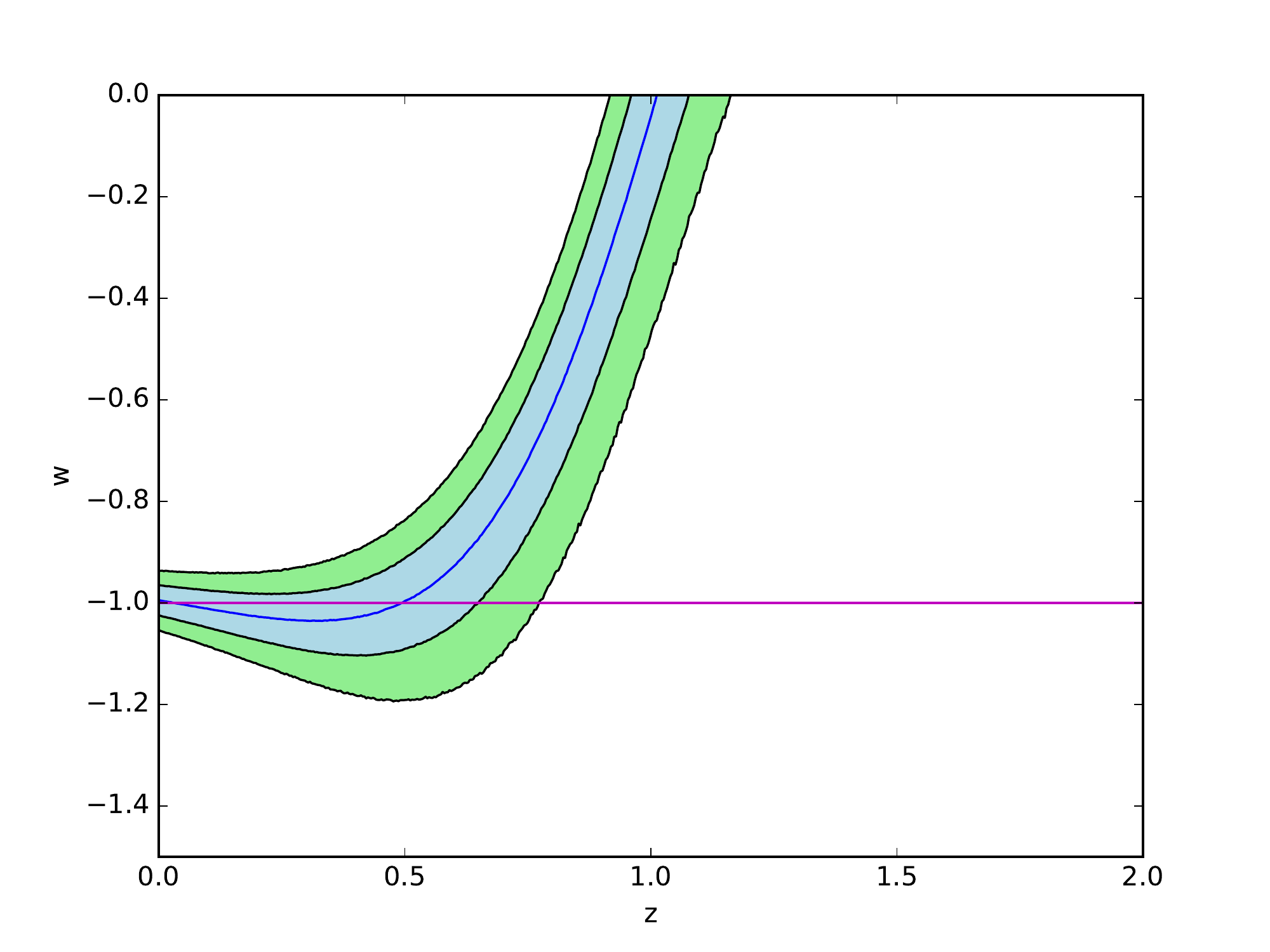}
\caption{The GP reconstructions of $D(z), D'(z)$, $D''(z)$ and the EoS of DE using Union 2.1 + H(z) + CMB + HII + GRB. The upper and lower panels from left to right correspond to the cases of $H_0=73.24\pm1.74$, $70$ and $66.93\pm0.62$ km s$^{-1}$ Mpc$^{-1}$. We have assumed $\Omega_{m0}=0.308\pm0.012$ and $\Omega_{k0}=0$.}\label{f7}
\end{figure}

To exhibit how much each probe is contributing better, we carry out both the GP reconstructions of $D(z), D'(z)$ and $D''(z)$, and those of the EoS of DE $\omega(z)$ utilizing the above five different probes. First of all, we consider the case of R16 $H_0=73.24\pm1.74$ km s$^{-1}$ Mpc$^{-1}$. From the upper left panels of both Fig. \ref{f2} and Fig. \ref{f3}, one can easily find that the reconstructions of $D(z), D'(z)$, $D''(z)$ and the EoS of DE are consistent with the $\Lambda$CDM model at the $2\sigma$ C.L.. However, using only 740 SNe Ia data points cannot characterize accurately the evolutional property of the EoS of DE. Then, we add the latest 30 H(z) data points and CMB shift parameter into the reconstruction processes. From the upper right panel of Fig. \ref{f2} and upper middle panel of Fig. \ref{f3}, we find that the H(z) probe gives a tighter constraint to $D(z), D'(z)$ and $D''(z)$ at redshifts $z\gtrsim0.42$, and that the EoS of DE is consistent with $\Lambda$CDM model at the $1\sigma$ C.L. when $z\gtrsim0.61$. This can be ascribed to the use of relatively high-$z$ H(z) data (see the right panel of Fig. \ref{f1}). From the middle left panel of Fig. \ref{f2} and upper right panel of Fig. \ref{f3}, one can find that CMB probe affects slightly the reconstructions of $D(z), D'(z)$, $D''(z)$ and the EoS of DE, since it just works well at extremely high redshift. Even if using a combination of 3 cosmological probes JLA + H(z) + CMB, we cannot still provide a more accurate and stricter constraint on the EoS of DE. Hence, we need the inputs of new data, i.e., more low-$z$ and high-$z$ data with high accuracy. In the middle right panel of Fig. \ref{f2} and lower left panel of Fig. \ref{f3}, supplying the 156 latest HII galaxy measurements, we find that the reconstruction of $D''(z)$ deviates much from the $\Lambda$CDM model at low redshifts, and that the EoS of the $\Lambda$CDM model lies out the $2\sigma$ confidence region when $z\in[0, 0.32]$ and $[0.82, 0.94]$. If supplying the 79 GRBs alone, we find that the reconstruction of $D''(z)$ deviates much from the $\Lambda$CDM model at relatively high redshifts at the $2\sigma$ C.L., and that that the EoS of the $\Lambda$CDM model lies out the $2\sigma$ confidence region when $z\in[0.11, 0.48]$ and $z\gtrsim1.09$. Furthermore, supplying both two probes, we find that the reconstruction of $D''(z)$ still deviates much from the $\Lambda$CDM model at relatively high redshifts at the $2\sigma$ C.L., and that the EoS of the $\Lambda$CDM model lies out the $2\sigma$ confidence region when $z\in[0, 0.54]$ and $z\gtrsim0.93$. One can easily conclude that the addition of HII galaxies and GRBs reduces apparently the uncertainties of the reconstructions, gives a quintom-like EoS of DE and implies that the DDE may actually exist in the late-time Universe.

In succession, since the value of $H_0$ affects the reconstruction results, we take the case of P16 $H_0=66.93\pm0.62$ km s$^{-1}$ Mpc$^{-1}$ into account. Using only JLA data, from the upper left panels of both Fig. \ref{f4} and Fig. \ref{f5}, we find that the reconstruction of $D'(z)$ deviates much from the $\Lambda$CDM model at low redshifts at the $2\sigma$ C.L., and that the EoS of DE are consistent with the $\Lambda$CDM model at the $2\sigma$ C.L.. Using a combination of JLA + H(z) or JLA + H(z) + CMB, one can find that the reconstructed $D(z)$ or $D'(z)$ deviates much from or is almost parallel to the $\Lambda$CDM model at the $2\sigma$ C.L., and that the EoS of DE with improved accuracy is still consistent with $\Lambda$CDM model at the $2\sigma$ C.L. However, when supplying the HII galaxies data, the reconstructed $D''(z)$ is consistent with the $\Lambda$CDM model at the $2\sigma$ C.L., which is different from the case of R16. Meanwhile, the EoS of the $\Lambda$CDM model lies out the $2\sigma$ confidence region when $z\in[0.77, 1.34]$. If supplying the GRB data, once again, the reconstructed $D''(z)$ deviates from the $\Lambda$CDM model at relatively high redshifts at the $2\sigma$ C.L., and the EoS of the $\Lambda$CDM model lies out the $2\sigma$ confidence region when only $z\gtrsim0.75$. In addition, supplying both two probes, we find that the EoS of the $\Lambda$CDM model lies out the $2\sigma$ confidence region when only $z\gtrsim0.76$. One can also conclude that the joint constraint from JLA + H(z) + CMB + HII + GRB still gives a quintom-like EoS of DE and indicates that the evolution of the late-time Universe may be actually dominated by the DDE.

In what follows, using a data combination of JLA + H(z) + CMB + HII + GRB, we also consider the cases of $H_0=60$, $70$ and $80$ km s$^{-1}$ Mpc$^{-1}$ in Fig. \ref{f6} and find that: (i) when $H_0=60$ km s$^{-1}$ Mpc$^{-1}$, the reconstructed $D(z), D'(z)$, $D''(z)$ and the EoS of DE deviate completely from the $\Lambda$CDM model over the $2\sigma$ C.L.; (ii) when $H_0=80$ km s$^{-1}$ Mpc$^{-1}$, the reconstructed $D(z), D'(z)$, $D''(z)$ deviate much from the $\Lambda$CDM model over the $2\sigma$ C.L., and the EoS of the $\Lambda$CDM model lies out the $2\sigma$ confidence region when $z\in[0, 0.69]$ and $z\gtrsim1.41$; (iii) when $H_0=70$ km s$^{-1}$ Mpc$^{-1}$, the reconstruction results are similar to the case of $H_0=73.24\pm1.74$ km s$^{-1}$ Mpc$^{-1}$. We conclude that too small and large $H_0$ values are disfavored by our GP reconstructions based on current cosmological observations. Since the JLA compilation has larger sample size and higher data quality than the Union 2.1 compilation, it is interesting and constructive to compare their reconstruction results with each other using the combined constraints from SNe Ia (JLA/Union 2.1) + H(z) + CMB + HII + GRB. In Fig. \ref{f7}, considering the cases of $H_0=73.24\pm1.74$, $70$ and $66.93\pm0.62$ km s$^{-1}$ Mpc$^{-1}$, we find that the JLA data just improves slightly the low-$z$ constraint on the EoS of DE. Based on the results in our previous work \cite{Deng}, we conclude that (i) five current cosmological probes support the P16's global measurement of $H_0$ very much in the low redshift range $z\in[0, 0.76]$ at the $2\sigma$ C.L. (see the lower right panel of Fig. \ref{f5}); (ii) the possible evidence of the DDE can be ascribed to the addition of HII galaxy and GRB data.

\section{Discussions and conclusions}
Since the accelerating Universe is discovered about two decades ago, one of the most urgent tasks in modern cosmology is to determine the evolution of the late-time Universe is actually dominated by the cosmological constant scenario or DDE. Previously \cite{Deng}, in the statistical framework of GP method, we have exhibited improved constraints on the EoS of DE by using the Union 2.1 SNe Ia data set, the 30 latest cosmic chronometer measurements and CMB data. However, the uncertainties of the constraints are very large in the low redshift range, and consequently we cannot give an tentative answer to this issue with high accuracy. In this follow-up study, we continue addressing this issue by using the JLA SNe Ia sample, the 30 latest cosmic chronometer data points, CMB data, the 156 latest HII galaxy measurements and 79 calibrated GRBs.

First of all, we review briefly on the GP methodology, describe the observed data and update the `` relations '' used in this analysis. Subsequently, we implement the GP reconstructions using the 5 above-mentioned different cosmological probes for the cases of both $H_0=73.24\pm1.74$ and $66.93\pm0.62$ km s$^{-1}$ Mpc$^{-1}$. We find that: (i) even if using a combination of JLA + H(z) + CMB, we cannot still provide a more accurate and stricter constraint on the EoS of DE; (ii) if only supplying HII galaxy or GRB data, the reconstructed EoS of DE is not always consistent with the $\Lambda$CDM model in the low redshift range at the $2\sigma$ C.L.; (iii) the joint constraints from JLA + H(z) + CMB + HII + GRB support the P16's global measurement of $H_0$ very much in the low redshift range $z\in[0, 0.76]$ at the $2\sigma$ C.L., give a quintom-like EoS of DE at the $2\sigma$ C.L. and imply that the evolution of the late-time Universe may be actually dominated by the DDE.

Furthermore, using a data combination of JLA + H(z) + CMB + HII + GRB, we also consider the cases of $H_0=60$, $70$ and $80$ km s$^{-1}$ Mpc$^{-1}$ and find that too small and large $H_0$ values are disfavored by our GP reconstructions based on current data (see Fig. \ref{f6}). Since the JLA compilation has larger sample size and higher data quality than the Union 2.1 compilation, we also compare their reconstruction results with each other using the combined constraints from SNe Ia (JLA/Union 2.1) + H(z) + CMB + HII + GRB. Considering the cases of $H_0=73.24\pm1.74$, $70$ and $66.93\pm0.62$ km s$^{-1}$ Mpc$^{-1}$, we find that the JLA data just improves slightly the low-$z$ constraint on the EoS of DE (see Fig. \ref{f7}).

It is worth noting that the possible evidence of the DDE can be ascribed to the addition of HII galaxy and GRB data. More specifically, 107 low-$z$ HII galaxy measurements lower the values of the reconstructed EoS of DE in the low redshift range, while 79 GRBs raise the values of the reconstructed EoS of DE in the relatively high redshift range. However, the quality of current data is still not enough good and shall be improved further. Therefore, we expect more and more high-quality data can give tighter constraints on the EoS of DE in the future.

\section{acknowledgements}
Deng Wang thanks Qi-Xiang Zou for programming and Lorenzo Zaninetti for very useful communications.

\end{document}